\documentclass[%
 reprint,
 amsmath,amssymb,
 aps,
]{revtex4-2}

\usepackage{graphicx,float}
\usepackage{dcolumn}
\usepackage{bm}
\usepackage{subfigure}
\usepackage{braket}
\usepackage{xcolor}
\usepackage{tabularx}
\usepackage{multirow}
\usepackage{amsmath}
\usepackage{diagbox}
\usepackage{makecell}
\usepackage{slashed}

\newcommand{\comment}[1]{{}}

\begin{document}

\preprint{APS/123-QED}

\title{Certifying Unknown Genuine Multipartite Entanglement by Neural Networks}

 \author{Zhenyu Chen$^{1}$}
 \author{Xiaodie Lin$^{1}$}
 \author{Zhaohui Wei$^{2,3,}$}\email{Email: weizhaohui@gmail.com}
 \affiliation{$^{1}$Institute for Interdisciplinary Information Sciences, Tsinghua University, Beijing 100084, China\\$^{2}$Yau Mathematical Sciences Center, Tsinghua University, Beijing 100084, China\\$^{3}$Yanqi Lake Beijing Institute of Mathematical Sciences and Applications, 101407, China
 }

 \begin{abstract}
    Suppose we have an unknown multipartite quantum state, how can we experimentally find out whether it is genuine multipartite entangled or not? Recall that even for a bipartite quantum state whose density matrix is known, it is already NP-Hard to determine whether it is entangled or not. Therefore, it is hard to efficiently solve the above problem generally. However, since genuine multipartite entanglement is such a fundamental concept that plays a crucial role in many-body physics and quantum information processing tasks, finding realistic approaches to certify genuine multipartite entanglement is undoubtedly necessary. In this work, we show that neural networks can provide a nice solution to this problem, where measurement statistics data produced by measuring involved quantum states with local measurement devices serve as input features of neural networks. By testing our models on many specific multipartite quantum states, we show that they can certify genuine multipartite entanglement very accurately, which even include some new results unknown before. We also exhibit a possible way to improve the efficiency of our models by reducing the size of features. Lastly, we show that our models enjoy remarkable robustness against flaws in measurement devices, implying that they are very experiment-friendly.
    \end{abstract}

    \maketitle

    \section{Introduction}
    Quantum entanglement plays a central role in many quantum information processing tasks, including quantum communication \cite{buhrman2010nonlocality}, quantum cryptography \cite{ekert1992quantum} and quantum key distribution \cite{scarani2009security}. As a consequence, certifying the existence of quantum entanglement is a fundamental problem.

    However, it has been proved that even for a bipartite quantum state that the density matrix is completely known, to determine whether it is entangled or not is already NP-Hard~\cite{gharibian2008strong}, implying that it is hard to generally solve this problem efficiently. Nevertheless, due to its importance, numerous approaches have been put forward to certify bipartite entanglement \cite{peres1996separability, horodecki1997separability, horodecki1996teleportation, guhne2009entanglement, doherty2004complete}.

    Meanwhile, we often face the situation that the target bipartite quantum state we would like to characterize is unknown to us, i.e., its density matrix is not available, making the task even tougher. For this, one may first reconstruct the density matrix by quantum state tomography~\cite{paris2004quantum,steffen2006measurement,poyatos1997complete}, and then try to solve the problem accordingly. However, it is well-known that this procedure is extremely expensive and can only be implementable when the size of the target quantum state is small. To overcome this difficulty, some realistic alternative approaches can be utilized, which include entanglement witness~\cite{bavaresco2018measurements,rosset2012imperfect,friis2019entanglement} and device-independent schemes~\cite{collins2002bell,moroder2013device,bowles2018device,xu2014implementation}. But they also suffer from other drawbacks, say being sensitive to operation errors or being easy to fail in providing valuable outcomes.

    When it comes to multipartite quantum states, the problem becomes even more complicated, as the mathematical structures of multipartite quantum entanglement are much richer than the bipartite case. Particularly, as a speical form of multipartite entanglement that is highly valuable, genuine multipartite entanglement has significant applications in quantum teleportation~\cite{yeo2006teleportation,chen2006general}, quantum state sharing~\cite{muralidharan2008perfect}, quantum metrology \cite{toth2012multipartite,hyllus2012fisher}, and even chemical and biological processes~\cite{sarovar2010quantum}. To certify it, many methods have been proposed. For example, when the full information of density matrices are known, quite a lot of mathematical criteria that can detect multipartite genuine entanglement have been proposed~\cite{guhne2010separability, szalay2015multipartite,ma2011measure,chen2012improved,de2011multipartite, huber2010detection, jungnitsch2011taming}.

    Similar to the bipartite case, we also have to handle the case that full information of target multipartite quantum states is not available, which is actually an extremely common and realistic problem from the viewpoint of engineering. For this, known methods such as quantum state tomography, entanglement witness, and device-independent schemes have been applied to certify genuine multipartite entanglement~\cite{bancal2011device,toth2005detecting, bourennane2004experimental,pal2011multisetting,moroder2013device,barreiro2013demonstration}. However, facing similar difficulties as in the bipartite case, it is not hard to understand that these approaches cannot provide satisfying solutions for this task. As a result, due to its central role in quantum computing and quantum engineerings, finding realistic approaches to certify unknown genuine multipartite entanglement is a challenging yet urgent task, which is also the main motivation of the current paper.

    Recently, machine learning approaches have been employed to characterize quantum properties \cite{girardin2022building,lu2018separability,chen2021detecting,chen2021entanglement,ren2019steerability,canabarro2019machine,lin2021quantifying,lin2022quantifying}. Different from analytic methods, machine learning is a data-driven approach which aims at making predictions on unseen data by learning from training data. In these works, following standard procedures of machine learning tasks, certain features are extracted from training quantum states, which are then fed into machine learning models. Then we train these models by adjusting the parameters they contain such that the models can make predictions on training quantum states with high accuracy, and if the model details are chosen properly, they can also make accurate predictions on target quantum states unseen before. Here, based on the chosen forms of features, different machine learning models can be designed to detect quantum properties.

    In this work, we exploit the possibility that utilizes neural networks to certify genuine multipartite entanglement for \emph{general} quantum states, where the features we choose is measurement statistics data produced by measuring involved quantum states with local measurement devices. Here each subsystem of involved multipartite quantum states is measured by at least two devices. Inspired by Bell experiments, we believe that this kind of measurement statistics data can reveal nontrivial quantum properties for target quantum states. 

    It turns out that our idea works very well. Particularly, we successfully train a series of neural network models that can certify unknown genuine multipartite entanglement very accurately, where the target quantum states are quite diverse. Taking 4-qubit quantum states for example, we first train a proper model, and then we run the same trained model on four different classes of 4-qubit quantum states, on each of which our model successfully certify genuine 4-qubit entanglement with accuracy over 99\%. Interestingly, our model even reports some new results that are unknown before, indicating that machine learning models can be highly valuable in such a challenging task. We confirm the high performance of neural networks on many other test quantum states, which include quantum state sets that are sampled randomly without any assumptions.

    Meanwhile, we also proposed a modified scheme called $k$-correlation to reduce the cost of our approach, and we show that in some cases where certain specific prior knowledge is available, the cost of our approach can be sharply reduced while the prediction accuracy is still comparable.

    Lastly, we provide evidence showing that our approach enjoys remarkable robustness against flaws in measurement devices, which implies that our approach is very experiment-friendly.

    \section{Deep learning and Entanglement structure}\label{setting}

    In this work, the certification of genuine multipartite entanglement is formulated as a supervised binary classification task, where the deep learning method is applied. Deep learning is a powerful machine learning model based on artificial neural networks. It has great representative ability and has been widely utilized in a variety of fields such as image recognition \cite{cireaan2012multi}, natural language processing \cite{gers2001lstm}, recommendation systems \cite{elkahky2015multi} and so on.

    For our task we apply fully connected neural networks (FNNs) to fit the training set, for more details one can see Refs.\cite{lecun1998gradient,albawi2017understanding}. Following the standard procedure of machine learning, we need to gather a training dataset and a test dataset, which have the form of $\left\{\left(\boldsymbol{x}_{1}, y_{1}\right), \ldots,\left(\boldsymbol{x}_{N}, y_{N}\right)\right\}$, where $N$ is the size of the set, $\boldsymbol{x}_{i}$ is the feature of the $i$-th sample, and $y_{i}$ is the label. The labels of training dataset are known to us, and the mission of the neural network is to predict the labels for the test dataset. When training the model, we input the features of training dataset into the model and adjusting its parameters such that it can produce correct labels for the training dataset. For this, a proper loss function, a reasonable configuration for the nerual network, and an efficient optimization method such as gradient-descent have to be chosen. If the model is trained properly, it can predict accurately the labels of test dataset unseen before.

    In this work we apply the deep learning method to certify genuine multipartite entanglement, which is also a typical binary-classification task. In general, a multipartite quantum state can involves many subsystems and thus its entanglement structure can be very complicated. An $n$-partite pure quantum state $\left|\Psi_{k-s e p}\right\rangle$ is called {\emph{$k$-separable}}, {where $1\leq k \leq n $,} {if and only if} it can be written as a {tensor} product of $k$ substates:
    \begin{eqnarray}\label{k-separable}
    \left|\Psi_{k-s e p}\right\rangle=\left|\Psi_{1}\right\rangle \otimes\left|\Psi_{2}\right\rangle \otimes \cdots \otimes\left|\Psi_{k}\right\rangle.
    \end{eqnarray}
    Correspondingly, a mixed state is called $k$-separable, if and only if it has a decomposition into $k$-separable pure states. A multipartite quantum state is called \emph{genuinely multipartite entangled} if it can not be written as $k$-separable for $k=2$, otherwise we call it \emph{biseparable state}. In addition, a quantum state is said to be of \emph{entanglement intactness $k$}, if it is $k$-separable but not $(k+1)$-separable.

    \section{Detecting Genuine Multipartite Entanglement for Qubit Systems}\label{Dectecting_qubits}

    \subsection{3-qubit {case}}\label{3-qubit}
    \subsubsection{The setup}

    As the simplest case, we first try to detect genuine multipartite entanglement for 3-qubit quantum states. As mentioned above, since our approach is based on a neural network, we need to prepare a large number of quantum states to train (and test) the neural network. In this work, we always sample random $d$-dimensional quantum state $\rho\in\mathcal{H}^d$ according to spectral decomposition
    \begin{eqnarray}\label{gen_state}
        \rho=\sum_{i=0}^{d-1}\lambda_i|u_i\rangle\langle u_i|.
        \end{eqnarray}
    Here we randomly choose nonnegative numbers $\lambda_i$'s such that they satisfy $\sum_i \lambda_i=1$. Then we generate a $d\times d$ Haar random unitary $U$, and set $|u_i\rangle$ to be the $i$-th column of $U$, which means that $\{|u_i\rangle\}$ forms a set of orthonormal basis for $\mathcal{H}^d$. Particularly, if $\lambda_i = 1$ for some $0\leq i\leq d-1$, $\rho$ will be a pure state. For now we need to sample 3-qubit quantum state, so we let $d=8$.

    After sampling 3-qubit quantum states, we need to prepare labels for them, i.e., indicating whether they are genuinely tripartite entangled or not. This will be done by the result given in Ref.\cite{guhne2010separability}, which proves that the following relation must be satisfied by any biseparable 3-qubit quantum states,
    \begin{equation}\label{sep_criteria}
     \begin{split}
    |\varrho_{2,3}|+
        &\left|\varrho_{2,5}\right|+\left|\varrho_{3,5}\right| \leqslant \sqrt{\varrho_{1,1} \varrho_{4,4}}+\sqrt{\varrho_{1,1} \varrho_{6,6}} \\
        &+\sqrt{\varrho_{1,1} \varrho_{7,7}}+\frac{1}{2}\left(\varrho_{2,2}+\varrho_{3,3}+\varrho_{5,5}\right),
     \end{split}
    \end{equation}
    where $\varrho_{i,j}$ is the ($i,j$)-th entry of the density matrix.

    Therefore, violating Eq.(\ref{sep_criteria}) implies the existence of genuine tripartite entanglement. Utilizing this criterion, we sample 30,000 genuinely tripartite entangled states. Concretely, we generate a 3-qubit state by Eq.(\ref{gen_state}), then substitute the density matrix into Eq.(\ref{sep_criteria}). If the equation.(\ref{sep_criteria}) does not hold, we keep this state, otherwise we drop it. Repeat this process until 30,000 genuinely tripartite entangled states are collected. We label each of them with `true' label.

    Next, we construct states with `false' labels by sampling 20,000 biseparable entangled quantum states and 20,000 fully separable states. When sampling biseparable entangled quantum states, all 3 possible partitions are included, that is, these quantum states are sampled according to the form
    \begin{equation}\label{depth_2}
        \begin{split}
        p_1 \rho_A\otimes \rho_{BC}+p_2 \rho_B\otimes \rho_{AC}+ p_3 \rho_C\otimes \rho_{AB}.
        \end{split}
       \end{equation}
    Here $p_1,p_2,p_3\geq 0 $ are randomly picked and satisfy $\sum_{i=1}^3p_i=1$. $\rho_{AB},\rho_{AC}, \rho_{BC}$ are bipartite entangled states whose partial transpositions are not positive semi-definite, implying the existence of partial entanglement~\cite{peres1996separability}.

    Putting the 30,000 genuinely tripartite entangled states, 20,000 biseparable entangled states, and 20,000 fully separable states sampled above together, we obtain the set of training quantum states, for which the correct labels have been pin down.

    Then we extract features for the training quantum states, which is essentially the measurement outcome statistics data when local quantum measurements are performed on each qubit. To choose measurement devices, we assign $M$ Haar random unitary matrices $U_{i1},U_{i2}, \cdots U_{iM} \in \mathbb{C}^{2 \otimes 2}$ to the $i$-qubit. Let $|u^k_{ij}\rangle$ be the $k$-th column of $U_{ij}$, then the $j$-th measurement device belonging to the $i$-th qubit can be expressed by the operators $P_{ij} = \{ |u^k_{ij}\rangle \langle u^k_{ij}|\}_{k=1}^{2}$, $j = 1, 2, \ldots M$. We emphasize that once the measurement devices are sampled, they remain unchanged during the whole process extracting features for all the training quantum states. What is more, when afterward we run the trained model on test quantum states, we also use the same measurement devices to extract features for the latter.

    According to Born's rule, the possibility that we obtain outcome $a_i\in\{1, 2\}$ when measuring the $i$-th qubit of $\rho$ with measurement device $P_{ix_i}$ is given by
    \begin{eqnarray}\label{Correlation}
    p\left(a_{1} a_{2} \ldots a_{n} \mid x_{1} x_{2} \ldots x_{n}\right)=\operatorname{Tr}\left(\left(\bigotimes_{i=1}^{n} P_{ix_{i}}^{a_{i}}\right) \rho\right),
    \end{eqnarray}
    where $x_i\in\{1, 2, \ldots M\}$. All the joint possibility distributions $p\left(a_{1} a_{2} \ldots a_{n} \mid x_{1} x_{2} \ldots x_{n}\right)$ combined are called a \emph{correlation}, which is the input features for our neural networks.

    In the 3-qubit case, we assign 2 measurement devices for each qubit, i.e., $M=2$, thus the dimension of the input features turns out to be $2^3\times 2^3=64$. 

    The FNN we employ in this case has one input layer, one sigmoid output layer, four hidden layers with $90,60,60,60$ neurons. To avoid overfitting, we add one drop out layer with a rate of $20\%$ after the first hidden layer. The loss function {adopted} for training is the binary crossentropy, which is widely used in binary classification tasks and can be written as:
    \begin{eqnarray}\label{Loss}
    \operatorname{loss}=\frac{1}{N}\left[-\sum_{i=1}^N y_i \log \hat{y}_i+\left(1-y_i\right) \log \left(1-\hat{y}_i\right)\right],
    \end{eqnarray}
    where $N$ is the number of the samples in the training set, $y_i$ is the {$i$-th} label of the training set, and $\hat{y}_i$ is the {$i$-th output} of the neural network. In the training process, the parameters of neural networks are adjusted constantly such that the loss function is optimized. In our problem, when $\hat{y}_i \geq 0.5$ the neural network outputs an outcome `true', meaning that the corresponding quantum state is genuinely tripartite entangled, otherwise an outcome `false', implying that the neural network predicts the corresponding quantum state to be biseparable.

    \subsubsection{Performance on GHZ-Werner states and W-Werner states}

    After training the neural network, we now test its performance in certifying genuine tripartite entanglement. First, we choose two well-known classes of quantum states, which are Greenberger-Horne-Zeilinger (GHZ) -Werner states
    \begin{eqnarray}\label{GHZ}
    \hat{\rho}_{\mathrm{GW}n} \equiv (1-p)\left|\mathrm{GHZ}_{n}\right\rangle\left\langle\mathrm{GHZ}_{n}\right|+p I/2^n,
    \end{eqnarray}
    and W-Werner states
    \begin{eqnarray}\label{W}
    \hat{\rho}_{\mathrm{WW}n} \equiv (1-p)\left|\mathrm{W}_{n}\right\rangle\left\langle\mathrm{W}_{n}\right|+p I/2^n,
    \end{eqnarray}
    where $|\mathrm{GHZ}_{n}\rangle=(|0\rangle^{\otimes n}+|1\rangle^{\otimes n})/\sqrt{2}$ and $|\mathrm{W}_{n}\rangle=(|100 \ldots 0\rangle+|010 \ldots 0\rangle+\ldots+|00 \ldots 01\rangle)/\sqrt{n}$. In our case, $n=3$.

    In fact, these two classes of quantum states have been well-studied~\cite{guhne2010separability,jungnitsch2011taming}. It turns out that for a GHZ-Werner state $\hat{\rho}_{\mathrm{GW}n}$, it is genuinely $n$-partite entangled if and only if $0 \leqslant p<1 /\left[2\left(1-2^{-n}\right)\right]$, and is biseparable when $p\geq1 /\left[2\left(1-2^{-n}\right)\right]$. For convenience, we call the value $1 /\left[2\left(1-2^{-n}\right)\right]$ the \emph{threshold} for the GHZ-Werner state to be genuinely $n$-partite entangled. When $n=3$, the threshold is $0.571$. Similarly, for a W-Werner state $\hat{\rho}_{\mathrm{WW}3}$, it is tripartite genuinely entangled if and only if $0\leqslant p<0.521$, thus the threshold is $0.521$. And the threshold of $\hat{\rho}_{\mathrm{WW}n}$ with $n \geq 4$ remains open.

    The above theoretical results provide us a very nice chance to examine the performance of the FNN model we have trained. For this, we first extract features for $\hat{\rho}_{\mathrm{GW}3}$ and $\hat{\rho}_{\mathrm{WW}3}$ by performing the measurement devices we sampled previously, where $p$ varies from 0 to 1 at intervals 0.001, and then we input these features into our neural network.

    It turns out that the prediction accuracy of our model on both of the two classes of test quantum states are over $99\%$. Particularly, the FNN model predicts that the $\hat{\rho}_{\mathrm{GW}3}$ and $\hat{\rho}_{\mathrm{WW}3}$ are genuinely entangled if and only if $p\leq 0.577$ and $p\leq 0.530$, respectively (recall that the exact thresholds for these two classes are 0.571 and 0.521 respectively). It is interesting to see that the output of the FNN model also has a threshold pattern, i.e., it always gives `true' predictions when $p$ is smaller than certain value, and always gives `false' otherwise. Our model only makes mistakes near the exact threshold for $\hat{\rho}_{\mathrm{GW}3}$ and $\hat{\rho}_{\mathrm{WW}3}$. Taking $\hat{\rho}_{\mathrm{GW}_3}$ for example, the neural network provides wrong predictions only when $p\in(0.572,0.577)$. 

    \subsubsection{Performance on random quantum states}

    In some sense, $\hat{\rho}_{\mathrm{GW}3}$ and $\hat{\rho}_{\mathrm{WW}3}$ are quite special 3-qubit quantum states. Therefore, to further assess the performance of the FNN model we have trained, we now test it on more general 3-qubit quantum states.

    In Ref.\cite{szalay2015multipartite}, the concept of tripartite entanglement of formation denoted $E_{f}(\rho)$ was proposed to quantify genuine tripartite entanglement, which is defined as
     \begin{eqnarray}\label{tripartite_E_f}
    E_{f}(\rho) \equiv \min _{\left|\psi_i\right\rangle}\left(\sum_i p_i \min \left\{S_i(A), S_i(B), S_i(C)\right\}\right),
    \end{eqnarray}
    where the first minimum is taken over all pure state decompositions of $\rho$ and $S_i(A), S_i(B), S_i(C) $ are the von Neumann entropies of the subsystem A, B and C of $|\psi_i\rangle$. It turns out that $E_{f}(\rho)$ equals zero for biseparable states \cite{szalay2015multipartite}, and can be lower bounded by $V_{\rho}$ defined as~\cite{schneeloch2020quantifying}
    \begin{eqnarray}\label{lower_bound}\scriptsize
    -S(A|BC)-S(B | A C)-S(C |A B)-2 \log \left(D_{\max }\right),
    \end{eqnarray}
    where $S(A|BC)$ is the von Neumann conditional entropy, and $D_{\max }$ is the maximum dimension of parties A, B, and C.

    $V_{\rho}$ is easy to compute when $\rho$ is given, and can also certify the existence of genuine tripartite entanglement if its value is positive. Combining this criterion with the approach that generates random quantum states given in Eq.(\ref{gen_state}), we sample 600 genuine tripartite entangled states. Together with 200 biseparable entangled states and 200 fully separable states, we construct a set of test quantum states. The features for these test quantum states are also generated by Eq.(\ref{Correlation}), where the chosen measurement devices are the same as those were utilized to prepare training dataset.

    We now run the neural network we have trained before on this test dataset. It turns out that our model again performs very well in this case, and the prediction accuracy is over $99\%$. Considering that the test dataset is generated randomly, it is fair to say that our model enjoys a very decent performance in the 3-qubit case.

    \subsection{$4$-qubit case}\label{4-qubit}
    
    Now we apply our approach on larger quantum systems, and turn to certify genuine 4-partite entanglement. Again, to train the neural network properly, we need to prepare a lot of training quantum states for which we know entanglement structure accurately. However, the criteria we utilized to certify genuine tripartite entanglement can not detect genuine 4-partite entanglement very well. Instead, we now apply a new criterion provided by Ref.\cite{ma2011measure}. Particularly, the concept of concurrence $C(\rho)$ is a genuine multipartite entanglement measure, and positive concurrence indicates the existence of genuine multipartite entanglement (see Ref.\cite{ma2011measure} for more details). Define
    \begin{eqnarray}
    \begin{split}
    \mathcal{F}(\rho, \psi)=& \sum_{1 \leqslant i \neq j \leqslant n} \sqrt{\left\langle\Psi_{i j}\left|\rho^{\otimes 2} \Pi\right| \Psi_{i j}\right\rangle} \\
    &-\sum_{1 \leqslant i \neq j \leqslant n} \sqrt{\left\langle\Psi_{i j}\left|\mathcal{P}_{i}^{\dagger} \rho^{\otimes 2} \mathcal{P}_{i}\right| \Psi_{i j}\right\rangle} \\
    &-(n-2) \sum_{1 \leqslant i \leqslant n} \sqrt{\left\langle\Psi_{i i}\left|\mathcal{P}_{i}^{\dagger} \rho^{\otimes 2} \mathcal{P}_{i}\right| \Psi_{i i}\right\rangle},\label{con_lower_bound}
    \end{split}
    \end{eqnarray}
    then it holds that $C(\rho)\geq\frac{1}{\sqrt{2}(n-1)}\mathcal{F}(\rho, \psi)$. Here $|\psi\rangle=\bigotimes_{i=1}^{n}\left|x_{i}\right\rangle=\left|x_{1} x_{2} \cdots x_{n}\right\rangle$ is an arbitrary product state {in} Hilbert space $\mathcal{H}=\mathcal{H}_{1} \otimes \mathcal{H}_{2} \otimes \cdots \otimes \mathcal{H}_{n}$. $\left|\psi_{i}\right\rangle=\left|x_{1} x_{2} \cdots x_{i-1} x_{i}^{\prime} x_{i+1} \cdots x_{n}\right\rangle$ and $\left|\psi_{j}\right\rangle=\mid x_{1} x_{2} \cdots$ $\left.x_{j-1} x_{j}^{\prime} x_{j+1} \cdots x_{n}\right\rangle$ {are} the product states obtained from $|\psi\rangle$ by applying  (independently) local unitaries to $\left|x_{i}\right\rangle \in \mathcal{H}_{i}$ and $\left|x_{j}\right\rangle \in \mathcal{H}_{j}$, respectively.  $\left|\Psi_{i j}\right\rangle:=\left|\psi_{i}\right\rangle\left|\psi_{j}\right\rangle$ is the tensor product of $\left|\psi_{i}\right\rangle$ and $\left|\psi_{j}\right\rangle$. $\Pi=\mathcal{P}_{1} \circ \mathcal{P}_{2} \circ \cdots \circ \mathcal{P}_{n}$, where $\mathcal{P}_{i}$ is the operator swapping the two copies of $\mathcal{H}_{i}$ in $\mathcal{H}^{\otimes 2}$, for each $i=1, \ldots, n$.
    
    We compute lower bounds for concurrence by randomly sampling a product state $|\psi\rangle$ and local Haar random unitaries. By this way, we sample $30,000$ states with positive $\mathcal{F}(\rho, \psi)$ to serve as `true' samples of our training set. Additional $20,000$ states labelled `false' are constructed for each entanglement intactness of $4,3,2$, where as before all possible partitions are included for each intactness. 
    
    After picking up the quantum states for training, we extract the features by the similar approach as before, that is, randomly sampling a set of local measurement devices and fixing them, and then collecting the outcome statistics data when measuring the training quantum states with these measurement devices, where the only difference is that we now have four parties, resulting in that the dimension of the input feature of the neural network is 256 accordingly.
    
    The neural network we employ here has a similar structure to the one used for the 3-qubit case, and the concrete configuration is listed in Table.\ref{table:FNN_configuration}. The loss function remains the binary crossentropy.
    
    \begin{table}[h] \scriptsize
        \caption{The configuration details of FNNs. In the first column, $n,d$ denotes the number and the dimension of the subsystems, respectively. }
        \setlength{\tabcolsep}{5mm}
        \renewcommand{\arraystretch}{1.2}
        {
        \begin{tabular}{lll}
            \hline
            \hline
            System {size} & Hidden layers \\
            \hline
            {$n=3,d=2$} & 90 , one drop out layer, 60, 60 , 60\\
            {$n=3,d=4$} & 20, one drop out layer, 12, 12,  12 \\
            {$n=4,d=2$} & 70, one drop out layer, 46, 46 , 46 \\
            {$n=5,d=2$}& 65, one drop out layer, 42, 42,  42 \\
            \hline
            \hline
        \end{tabular}}
        \label{table:FNN_configuration}
        \end{table}
    
    After training the neural network, we use known analytical results on genuine 4-partite entanglement to evaluate the performance of our model. For this, we focus on four kinds of 4-qubit quantum states mixed with white noise, which are $|GHZ_{4}\rangle$,
    $|W_{4}\rangle$, $\left|Cl_{4}\right\rangle$ and $\left|D_{24}\right\rangle$. The definition of $|G H Z_{4}\rangle$ and $|W_{4}\rangle$ have been given in Eq.(\ref{GHZ}) and Eq.(\ref{W}), and $\left|C l_{4}\right\rangle$ and $\left|D_{24}\right\rangle$ are represented as
    \begin{align}
    	\left|C l_{4}\right\rangle=&(|0000\rangle+|0011\rangle+|1100\rangle-|1111\rangle) / 2, \\
    	\left|D_{24}\right\rangle=&\frac{1}{\sqrt{6}}(|0011\rangle+|1100\rangle+|0101\rangle+|0110\rangle+|1001\rangle. \nonumber\\
        &+|1010\rangle)
    \end{align}
    
    The exact thresholds for $|GHZ_{4}\rangle$ and $\left|Cl_{4}\right\rangle$ with white noise have been found to be {0.533} and 0.614 respectively~\cite{jungnitsch2011taming}. That is to say, $(1-p)\ket{GHZ_{4}}\bra{GHZ_{4}}+pI/16$ is genuine 4-partite entangled if and only if $0\leq p<0.533$, and similar for $\ket{Cl_{4}}$. However, to the best of our best knowledge the exact thresholds for $|W_{4}\rangle$ and $\left|D_{24}\right\rangle$ mixed with white noise are unknown. And the best-known results for $|W_{4}\rangle$ and $\left|D_{24}\right\rangle$ mixed with white noise to be genuine 4-partite entangled is that $p<0.526$ and $p<0.539$ respectively~\cite{jungnitsch2011taming}.
    
    Similar to the 3-qubit case, we gather one test set of quantum states for each of the above four classes of 4-qubit quantum states, where $p$ varies at intervals of 0.001. Again, the features for these test quantum states are extracted by the same measurement devices as what we utilized in training the neural network. The labels for these test quantum states are largely known from the analytical results given in Ref.\cite{jungnitsch2011taming}, and we use them to benchmark our model. 
    
    On the four test dataset, the results provided by our neural network are listed in Table.\ref{table:results_4,5 qubits}. It turns out that on all the instances that analytical results are known ($0\leq p<1$ for $|GHZ_{4}\rangle$ and $\left|Cl_{4}\right\rangle$, $0\leq p<0.526$ for $|W_{4}\rangle$, and $0\leq p<0.539$ for $\left|D_{24}\right\rangle$), overall our model makes excellent predictions. Interestingly, the results given by the neural network again have a threshold pattern, which reveals that the thresholds for the $|W_{4}\rangle$ class and the $\left|D_{24}\right\rangle$ class are possibly around $0.635$ and $0.585$ respectively, indicating the potential value of our neural network again, particularly in quantum engineering areas. 
    
    \begin{table}[h] \scriptsize
        \caption{Performance of the nerual network for the 4-qubit case. For the classes marked by~*, the exact threshold has been known. The last two columns denote the thresholds given by our model and the best-known noise tolerance for the existence of genuine multipartite entanglement.}
        \setlength{\tabcolsep}{1.5mm}
        \renewcommand{\arraystretch}{1.2}
        {
        \begin{tabular}{cccc}
            \hline
            \hline
            States & Accuracy & Threshold (FNN) &  Best-known \\
            \hline
            $|G H Z_{4}\rangle ^*$ & $99\%$ &0.523 &0.533 \cite{jungnitsch2011taming}\\
            $|W_{4}\rangle$ & $100\%$ & 0.635 &0.526 \cite{jungnitsch2011taming}\\
            $\left|C l_{4}\right\rangle^*$ & $99.7\%$ &0.611 &0.614 \cite{jungnitsch2011taming}\\
            $|D_{2,4}\rangle$ & $100\%$ &0.585 &0.539 \cite{jungnitsch2011taming}\\
            Random  & $99.4\%$ & / & / \\
            \hline
            \hline
        \end{tabular}}
        \label{table:results_4,5 qubits}
        \end{table}
    
    To provide more convincing evidence that the neural network we trained works well, we now test it on more general 4-qubit quantum states. It has been proven in Ref.\cite{de2011multipartite} that for a 4-qubit quantum state $\rho$, if one of the inequalities
    \begin{eqnarray}
        \begin{split}
            \left\|M_{22}\left(T_{i_1 i_2 i_3 i_4}\right)\right\|_k> \begin{cases}2 \sqrt{k} & 1 \leq k \leq 3 \\ 1+2 k / 3 & 4 \leq k \leq 9\end{cases}\label{tensor_criterion}
        \end{split}
        \end{eqnarray}
    holds, then the state must be genuine multipartite entangled. Here $T_{i_1 i_2 i_3 i_4} = {\rm Tr}(\sigma_{i_1} \otimes \cdots \otimes \sigma_{i_4} \rho)$, {$i_j=1,2,3$ for each $j=1,\cdots,4$} and $\sigma_1, \sigma_2, \sigma_3$ are the pauli matrix $X, Y, Z.$ $\left\|M_{22}\left(T_{i_1 i_2 i_3 i_4}\right)\right\|_k=$ $\left(\left\|T_{\underline{i_1} \underline{i_2} i_3 i_4}\right\|_k+\left\|T_{\underline{i_1} i_2 i_3 i_4}\right\|_k+\left\|T_{\underline{i_1} i_2 i_3 \underline{i_4}}\right\|_k\right) / 3$, $T_{\underline{i_1} i_2 \underline{i_3} i_4}=\sum_{i_1 \cdots i_4} T_{i_1 i_2 i_3 i_4}\left|i_1 i_3\right\rangle\left\langle i_2 i_4\right|$, and $\left\| \cdot \right\|_k $ is the Ky Fan k norms defined as the first $k$ largest singular values of the matrix~\cite{horn2012matrix}.
    
    Similar as before, we sample 600 genuine 4-partite entangled states by first randomly generating quantum states by Eq.(\ref{gen_state}), and then picking up those satisfying Eq.(\ref{tensor_criterion}). Together with 600 biseparable states, we construct a nice test set of quantum states. The experimental result is also listed in Table.\ref{table:results_4,5 qubits}. As we can see, the prediction accuracy is over $99\%$, indicating a very decent and stable performance of our model.

     \section{Detecting Genuine Multipartite Entanglement for Qudit States}\label{Dectecting_general_GME}
    
     Next we turn to more general multipartite quantum states, where each subsystem can be higher-dimensional, and the number of subsystems can also be larger.
    
     We would like to point out that in this case training neural networks is more challenging. The reason is that after randomly sampling candidate quantum states for training, it is hard to provide correct labels for them. For example, if we apply the criteria that we have utilized previously, say Eq.(\ref{lower_bound}) and Eq.(\ref{con_lower_bound}), on $5$-qubit quantum states or higher dimensional quantum states, it is very difficult to certify the existence of genuine multipartite entanglement. In fact, we also tried to use the technique introduced in Ref.\cite{jungnitsch2011taming} for the same task, which is actually an optimization approach based on semi-definite programming, and we find that applying this method is very time-consuming.
    
    After careful comparisons, we realize that the approach proposed in Ref.\cite{shen2020construction} is very suitable to provide labels for candidate training quantum states. Basically, this approach points out that an ($n+2$)-partite {genuinely} entangled state can be obtained by merging two ($n+1$)-partite {genuinely} entangled states. Specifically, suppose $\rho_{A C_{1,1} C_{1,2} \cdots C_{1, n}}$ and $\tau_{B C_{2,1} C_{2,2} \cdots C_{2, n}}$ are two $(n+1)$-partite {genuinely} entangled states in the Hilbert spaces $\mathcal{H}_{A C_{1,1} C_{1,2} \cdots C_{1, n}}$ and $\mathcal{H}_{B C_{2,1} C_{2,2} \cdots C_{2, n}}$ respectively. Then an $(n+2)$-partite genuinely entangled state can be constructed by applying the operater $\otimes_{K_c}$ as follows
    \begin{eqnarray}
        \begin{split}
            \mathcal{H}_{A C_{1,1} C_{1,2} \cdots C_{1, n}} \otimes_{K_c} \mathcal{H}_{B C_{2,1} C_{2,2} \cdots C_{2, n}}:= \\
            \mathcal{H}_A \otimes \mathcal{H}_B \otimes(\mathcal{H}_{C_{1} C_{2} \cdots C_ n}),
            \label{construct_GME}
        \end{split}
        \end{eqnarray}
    where $\mathcal{H}_{C_{j}} = \mathcal{H}_{C_{j,1}} \otimes \mathcal{H}_{C_{j,2}}$, $ 1 \leq j\leq n$. By definition, $\rho_{A C_{1,1} C_{1,2} \cdots C_{1, n}} \otimes_{K_c} \tau_{B C_{2,1} C_{2,2} \cdots C_{2, n}}$ is an $(n+2)$-partite state of systems $A, B$, and $C_j$ 's, {where} $C_j:=\left(C_{1, j} C_{2, j}\right)$ {for} $1 \leqslant j \leqslant n$.
    
    By this method, we can use the small-scale genuine multipartite entangled states sampled previously to build larger genuine multipartite entangled states in an iterative manner, which in turn are utilized to train neural networks customized for large quantum systems. 
    
    Meanwhile, note that the entanglement dimension is enlarged after the merging operation introduced above. For example, if we have two bipartite entangled quantum states in Hilbert space $\mathcal{H}^{4}\otimes\mathcal{H}^{2}$, after the merging operation we can obtain a genuine multipartite entangled state in $\mathcal{H}^{4}\otimes\mathcal{H}^{4}\otimes\mathcal{H}^{4}$. 
    
    \subsection{Tripartite qudit case}\label{higher_dimension}
    
    We first apply our approach on tripartite qudit systems. To sample genuine tripartite entangled quantum states for training, we first generate a lot of bipartite entangled states, which can be achieved by the positive partial transpose criterion. Then with the merging operation introduced in Ref.\cite{shen2020construction} to generate many desirable genuine tripartite entangled state. In this way, we eventually collect 20,000 genuine tripartite entangled states in $\mathcal{H}^4\otimes \mathcal{H}^4 \otimes \mathcal{H}^4$. Combining them with 40,000 biseparable states in the same Hilbert space, we finish preparing the set of training quantum states.
    
    Next we extract features for the training quantum states, which is again the measurement statistics data produced by measuring these quantum states locally. Since the dimensions of subsystems are now larger, we increase the number of measurement devices for each party from two to three. Accordingly, we let $M=3$ in Eq.(\ref{Correlation}), and the feature size of the neural network becomes $576$. After extracting the features of training quantum states, we input them into the neural network to train our model. The configuration of the FNN for this case is also listed in Table. \ref{table:FNN_configuration}.
    
    To test the model we just trained, we first consider the generalized GHZ state $\left|GHZ_{4 3}\right\rangle=\frac{1}{2} \sum_{i=0}^{3}|i\rangle^{\otimes 3}$ with white noise:
    $$
    \rho=p\left|GHZ_{4 3}\right\rangle\left\langle GHZ_{4 3}\right|+(1-p) \frac{1}{4^3} I.
    $$
    By running our model on this class of quantum states, we observe that it makes the `true' prediction when $p>0.161$, while it has been proved analytically that this class of quantum states is genuine tripartite entangled when $p>0.157$~\cite{huber2010detection}.
    
    As the second class of test quantum states, we now turn to the qutrit state
    $$
    \rho=\frac{1-\alpha-\beta}{27} I+\frac{\alpha}{3} \rho_{\text {bisep }}+\beta \left|GHZ_{33}\right\rangle\left\langle GHZ_{33}\right|,
    $$
    where $\rho_{\text {bisep }}=$ $|0\rangle\langle 0|\otimes(|00\rangle+|11\rangle+|22\rangle)(\langle 00|+\langle 11|+\langle 22|)$. To employ our model for these qutrit states, we embed each qutrit into a 4-dimensional Hilbert space by padding zeros to the redundant bases to make these test quantum states have the same dimension with the training quantum states and the measurement devices that extract features.
    
\begin{figure}[!ht]
    	\centering
    	\includegraphics[width=0.35\textwidth]{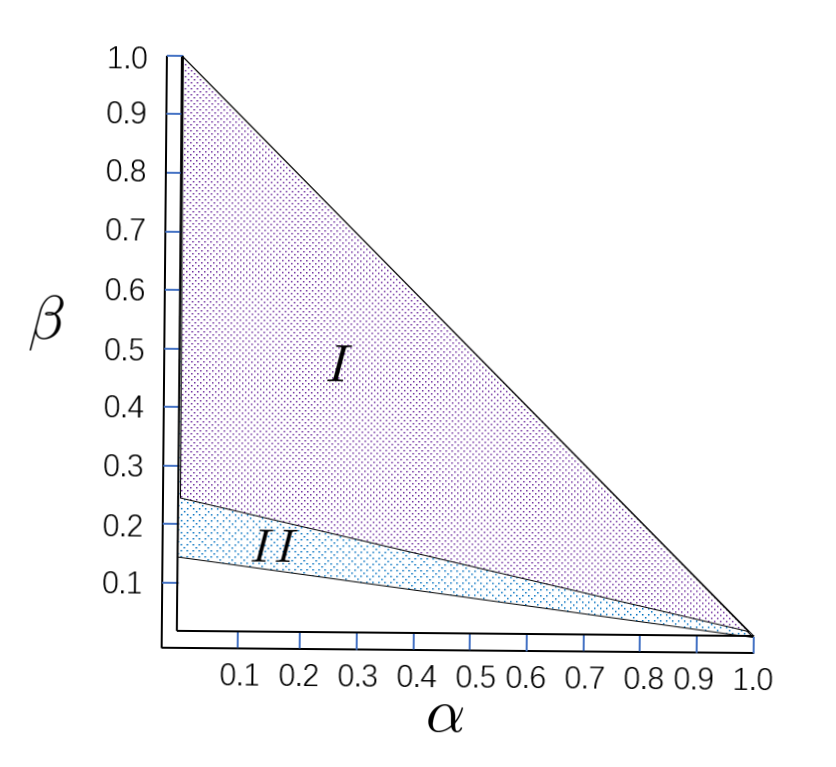}
    	\caption{{Parameter space for genuine tripartite entanglement certified by different methods.} Area $\uppercase\expandafter{\romannumeral1}$ is given in {Ref.}\cite{huber2010detection}. The neural network certifies both Area $\uppercase\expandafter{\romannumeral1}$ and Area $\uppercase\expandafter{\romannumeral2}$. The $\beta$-coordinate on the lower boundary of Area $\uppercase\expandafter{\romannumeral2}$ equals 0.14 when $\alpha = 0$. }
    	\label{fig:result_ghz_sep}
    \end{figure}

    The result given by the neural network is depicted in Fig.\ref{fig:result_ghz_sep}. Particularly, Area I is the parameter space that has been proved to be genuine tripartite entangled in Ref.\cite{huber2010detection}, which is strictly smaller than the parameter space that the neural network reports to be genuine tripartite entangled, as the latter also contain Area II.
    
    To double check the correctness of the results provided by the neural network, we perform independent numerical calculations to study the case that $\alpha=0$. By brute-force search for all possible biseparable decompositions of $\rho$, we find that $\rho$ is genuine tripartite entangled only when $\beta$ is larger than 0.13, which is very close to the result given by our model. We believe that the neural network provides very reliable results in this case.
    
    \subsection{$5$-partite qudit case}\label{n_qubit_case}
    
    As the last demonstration of our approach, we now use it to detect genuine 5-qudit entanglement. Based on the 4-qubit genuinely entangled states sampled in Sec.\ref{4-qubit}, we use the merging technique introduced in Ref.\cite{shen2020construction} to generate 20,000 5-partite genuinely entangled states in $(\mathcal{H}^{2})^{\otimes 2}\otimes (\mathcal{H}^{4})^{\otimes 3}$. To unify the dimension of subsystems for simplicity, we embed these quantum states into a larger Hilbert space $(\mathcal{H}^4)^{\otimes 5}$ also by padding zeros to the redundant bases. Together with 60,000 biseparable states with `false' labels, we obtain the whole set of training quantum states. Then as usual, we extract features for these states by measuring them with local measurement devices, where each qudit has three devices. The configuration of the neural network we employ in this case can be found in Table.\ref{table:FNN_configuration}.
    
    We now test the performance of this trained model. Due to that quantum states which have been certified analytically to be genuine 5-qudit entanglement are rare, we first test our model with the $|GHZ\rangle_5$ with white noise, which are accordingly embedded into the Hilbert space $(\mathcal{H}^4)^{\otimes 5}$. Again, when testing we vary the value of $p$ at intervals 0.001. Recall that the 5-qubit GHZ Werner state has been proved to be genuinely entangled if and only if $0 \leqslant p<0.516$, and it turns out that our model achieves over $98\%$ prediction accuracy on this class of quantum states.
    
    In order to further test the power of our model, we construct another test set composed of more general quantum states. For this, we sample 600 biseparable states labeled `false', and 600 genuine 5-qudit entangled states labeled `true' with the technique introduced in Eq.(\ref{construct_GME}). Combining these 1200 quantum states together, we obtain the new test set, on which our model certify genuine multipartite entanglement with accuracy over 99\%. 
    
    With the successes of all the models we have trained above, we expect that if one extends the neural network approach to more complicated cases of target quantum states, the performance will also be decent.
    
    \section{Reducing the cost of quantum measurements}
    
    We have shown the great potential of detecting genuine multipartite entanglement by machine learning methods, where the features are measurement statistics data. However, if the number of subsystems goes up, the cost of quantum measurements (or the size of features) will increase very quickly. With this in mind, we now propose the so-called \emph{$k$-correlation} scheme to reduce the cost of quantum measurements.
    
    \subsection{$k$-correlation {scheme}}\label{Gen_kcorr}
    
    Recall that when extracting feature for training and test quantum states, a set of local quantum measurements are performed on these quantum states to produce correlation data, where many cross terms are involved. For example, suppose Alice (Bob) possesses measurement devices $A_1, A_2$ ($B_1, B_2$), and she (he) measures the subsystem of a bipartite quantum state locally. The correlation data they produce comes from four different combinations of measurement devices, i.e., $A_1B_1$, $A_1B_2$, $A_2B_1$ and $A_2B_2$, where $A_1B_2$, $A_2B_1$ are called the cross terms. If we ignore these two cross terms, we can reduce the size of correlation data by half in this case. By generalizing this idea to $n$-partite quantum state $\rho$, we define the concept of $k$-correlation data, which is given by
    \begin{eqnarray}\label{k-Correlation}
        p\left(a_{1} a_{2} \ldots a_{n} \mid x\right)=\operatorname{Tr}\left(\left(\bigotimes_{i=1}^{n} P_{ix}^{a_{i}}\right) \rho\right),
        \end{eqnarray}
    where $a_i\in[{d_i}] \equiv\{1, \ldots, d_i\}$ represents the outcome for measurement device $P_{ix}$, {here $x\in\{1, 2\ldots k\}$}. In this section, we will try to certify genuine multipartite entanglement by neural networks with features being $k$-correlation data, and for convenience, we call this approach \emph{the $k$-correlation scheme}.
    
    \subsection{Detecting genuine 4-qubit entanglement using {the $k$-correlation scheme}}\label{4qubit_kcorr}
    
    Since analytical results on genuine 4-qubit entanglement are relatively rich, we now test the performance of the $k$-correlation scheme in this case. To be specific, we will reuse the same sets of training and test quantum states as before, and the only difference between the two models is the forms of features, which involves both the training and test stages.
    
    After retraining our model based on 5-correlation data, we again test the model with the four classes of 4-qubit quantum states, i.e., $|GHZ_{4}\rangle$, $|W_{4}\rangle$, $|Cl_{4}\rangle$ and $|D_{24}\rangle$ mixed with white noise. The results on prediction accuracy for each class are listed in Table.\ref{table:results_4qubit-5corr}.
    
    \begin{table}[h]
        \caption{Performance of {the neural network based on $5$-correlation data.}}
        \setlength{\tabcolsep}{5mm}{
        \begin{tabular}{lll}
            \hline
            \hline
            States & accuracy  \\
            \hline
            $|G H Z_{4}\rangle ^*$ & $86\%$ \\
            $|W_{4}\rangle$ & $100\%$ \\
            $\left|C l_{4}\right\rangle^*$ & $99.6\%$ \\
            $|D_{2,4}\rangle$ & $100\%$ \\
            \hline
            \hline
        \end{tabular}}
        \label{table:results_4qubit-5corr}
        \end{table}
    
    As we can see, the performance of the new model is worse than the original one based on full correlation data. For example, the prediction accuracy for the GHZ class is lower compared with the old result. To extract more information from the training and test quantum states, we offer each party more measurement devices, and generate $k$-correlations with $k\ge6$. It turns out that even if we let $k=16$, which means the features now have the same size as the original model, the new model still can not certify genuine 4-qubit entanglement for the GHZ-Werner class with accuracy over $90\%$. This reveals an interesting fact that cross terms in correlation data is very crucial to extract quantum properties for underlying quantum states.
    
    \subsection{Detecting genuine multipartite entanglement for graph states}\label{graph_states}
    
    Though the power of the $k$-correlation scheme is weaker than the original scheme based on full correlation data, we now show that if certain prior knowledge on target quantum states are known, the $k$-correlation scheme still enjoys a nice performance.
    
    To demonstrate this fact, in this subsection we utilize the $k$-correlation scheme to predict whether graph states are genuine multipartite entangled. It is well-known that graph states have significant applications in quantum computing and quantum information \cite{hein2004multiparty,hein2006entanglement,keet2010quantum,looi2008quantum}.
    
    The motivation to apply the $k$-correlation scheme on graph states is very clear. Suppose we have a graph state consisting of $12$ qubits, then if we let full correlation data serve as features for the neural network, the feature size will be $2^{24}$ even if each qubit is measured by only two measurement devices. As a comparison, if we apply the $k$-correlation scheme, the feature size will be $2^{12}\cdot k$, which can be much smaller. Interestingly, it turns out that the $k$-correlation scheme works well in this case.
    
    Specifically, we first randomly generate 600 10- (11-, 12-) qubit graph states to train the neural network, where the corresponding labels (whether the graph states are genuine multipartite entangled) can be determined by the connectivity of the underlying graphs \cite{west2001introduction}. Here the FNN model we choose has 3 hidden layers that contains 50, 25, 10, neurons respectively. After training, we test the model by running it on another 200 random graph states. The result is that even $4$-correlation data is already enough to achieve a prediction accuracy of $97\%$.

    It can be seen that the cost of quantum measurement in the $k$-correlation scheme has been sharply reduced, where the number of measurement device combinations goes down dramatically. This indicates clearly that prior knowledge can be very helpful in our approach. In fact, similar result has also reported in Ref.\cite{chen2021entanglement}, where a machine learning method is applied to certify the entanglement structure of pure and noisy generalized GHZ states based on expectation values of local observables, and a very nice performance is achieved due to the fact that the training and test quantum states there have very similar configuration, which is essentially a kind of prior knowledge. We stress that in the current work our models can handle much more general quantum states, and as we have pointed out, test quantum states we choose are often sampled randomly without any assumptions.
    
    We remark that the $k$-correlation scheme and the original scheme based on full correlation data are two extreme ends to apply our neural network approach. From the viewpoint of quantum engineering, we believe that there are many other ways to exploit correlation data with neural network, and a proper tradeoff between efficiency and prediction accuracy can be found out based on specific application scenarios.
    
    \section{Robustness of our models}
    
    Recall that in each model we have trained, at the beginning we always randomly sample a set of measurement devices and fix them, then the features of all training and test quantum states are extracted based on these devices. Naturally, one may ask, if we repeat the whole process and sample a different set of measurement devices, can we obtain better prediction performance?
    
    Interestingly, in all the models we have trained, we do make this kind of comparisons, and it turns out that measurement devices generated by different choices of Haar random matrices can only change the prediction performance very slightly.
    
    Particularly, in the 4-qubit case we also try to extract features with the observables taken from Ref.\cite{lu2018entanglement}, which is widely used to witness entanglement intactness. Specifically, we assign each party the following two observables
    \begin{eqnarray}\label{optimal_measure}
        &A_{1}=\sigma_z, \\
        &A_{2}=A_+,
    \end{eqnarray}
    where $\sigma_z$ is the Pauli matrix Z and $A_{+}=\cos \left(\frac{n+1}{2 n} \phi\right) \sigma_x+\sin \left(\frac{n+1}{2 n} \phi\right) \sigma_y$. In our case, $n=4$ and $\phi = 1.1055$. After repeating the training and test procedures with the new way to extract features, we observe that the accuracy of detecting genuine 4-qubit entanglement keeps almost unchanged.
    
    To provide further evidence for the above robustness, we also introduce the mechanism of `learnable measurement' to find out what measurement devices are optimal in our models. For this, we parameterize involved local measurement devices, and search for the optimal family of parameters by optimization approaches~\cite{lin2022quantifying}. Again, the result shows that even the optimized measurement devices still provide almost the same prediction accuracy.
    
    We would like to stress that from the point of view of engineering, this kind of robustness is very valuable. This implies that, even if the quality of the measurement devices we utilize is poor, as long as their workings keep stable, our approach can still works very well.
    
    \section{Conclusion}
    
    To summarize, in this work we demonstrate that neural networks can certify genuine multipartite entanglement very accurately based on the measurement statistics data produced by measuring involved quantum states with local measurements. Specifically, we successfully train neural networks to detect genuine multipartite entanglement for many multipartite qubit and qudit systems. By testing these trained models on various multipartite quantum states, we observe that the prediction accuracy is very high. Particularly, in many cases the test quantum states we choose are sampled randomly, which provides convincing evidence showing that neural networks can work very well in such a challenging task.
    
    In addition, to improve the efficiency of our models, we propose the $k$-correlation scheme to reduce the cost of quantum measurements. We show that when prior knowledge on target quantum states is known, the performance of this scheme can still be very good. This indicates that one can adjust model details to save computational resources according to application scenarios.
    
    Meanwhile, we remark that our approach can also be combined with the idea of shadow tomography~\cite{huang2020predicting}, by which one can generate correlation data based on classical shadow of involved quantum states, and thus reduce the cost of quantum operations dramatically. 
    
    Lastly, we show that our models enjoy remarkable robustness against flaws in involved quantum measurements, and this implies that our models are very experiment-friendly. We expect that in future our neural network approach can be applied to certify unknown genuine multipartite entanglement, providing a realistic solution to this fundamental yet difficult problem.
\begin{acknowledgements}
We thank Zhangqi Yin for helpful comments. This work was supported by the National Key R\&D Program of China, Grants No. 2018YFA0306703, 2021YFE0113100, and the National Natural Science Foundation of China, Grant No. 61832015, 62272259.
\end{acknowledgements}

\bibliography{ref}

\begin{thebibliography}{60}%
\makeatletter
\providecommand \@ifxundefined [1]{%
 \@ifx{#1\undefined}
}%
\providecommand \@ifnum [1]{%
 \ifnum #1\expandafter \@firstoftwo
 \else \expandafter \@secondoftwo
 \fi
}%
\providecommand \@ifx [1]{%
 \ifx #1\expandafter \@firstoftwo
 \else \expandafter \@secondoftwo
 \fi
}%
\providecommand \natexlab [1]{#1}%
\providecommand \enquote  [1]{``#1''}%
\providecommand \bibnamefont  [1]{#1}%
\providecommand \bibfnamefont [1]{#1}%
\providecommand \citenamefont [1]{#1}%
\providecommand \href@noop [0]{\@secondoftwo}%
\providecommand \href [0]{\begingroup \@sanitize@url \@href}%
\providecommand \@href[1]{\@@startlink{#1}\@@href}%
\providecommand \@@href[1]{\endgroup#1\@@endlink}%
\providecommand \@sanitize@url [0]{\catcode `\\12\catcode `\$12\catcode
  `\&12\catcode `\#12\catcode `\^12\catcode `\_12\catcode `\%12\relax}%
\providecommand \@@startlink[1]{}%
\providecommand \@@endlink[0]{}%
\providecommand \url  [0]{\begingroup\@sanitize@url \@url }%
\providecommand \@url [1]{\endgroup\@href {#1}{\urlprefix }}%
\providecommand \urlprefix  [0]{URL }%
\providecommand \Eprint [0]{\href }%
\providecommand \doibase [0]{https://doi.org/}%
\providecommand \selectlanguage [0]{\@gobble}%
\providecommand \bibinfo  [0]{\@secondoftwo}%
\providecommand \bibfield  [0]{\@secondoftwo}%
\providecommand \translation [1]{[#1]}%
\providecommand \BibitemOpen [0]{}%
\providecommand \bibitemStop [0]{}%
\providecommand \bibitemNoStop [0]{.\EOS\space}%
\providecommand \EOS [0]{\spacefactor3000\relax}%
\providecommand \BibitemShut  [1]{\csname bibitem#1\endcsname}%
\let\auto@bib@innerbib\@empty
\bibitem [{\citenamefont {Buhrman}\ \emph {et~al.}(2010)\citenamefont
  {Buhrman}, \citenamefont {Cleve}, \citenamefont {Massar},\ and\ \citenamefont
  {De~Wolf}}]{buhrman2010nonlocality}%
  \BibitemOpen
  \bibfield  {author} {\bibinfo {author} {\bibfnamefont {H.}~\bibnamefont
  {Buhrman}}, \bibinfo {author} {\bibfnamefont {R.}~\bibnamefont {Cleve}},
  \bibinfo {author} {\bibfnamefont {S.}~\bibnamefont {Massar}},\ and\ \bibinfo
  {author} {\bibfnamefont {R.}~\bibnamefont {De~Wolf}},\ }\bibfield  {title}
  {\bibinfo {title} {Nonlocality and communication complexity},\ }\href@noop {}
  {\bibfield  {journal} {\bibinfo  {journal} {Reviews of modern physics}\
  }\textbf {\bibinfo {volume} {82}},\ \bibinfo {pages} {665} (\bibinfo {year}
  {2010})}\BibitemShut {NoStop}%
\bibitem [{\citenamefont {Ekert}(1992)}]{ekert1992quantum}%
  \BibitemOpen
  \bibfield  {author} {\bibinfo {author} {\bibfnamefont {A.~K.}\ \bibnamefont
  {Ekert}},\ }\bibfield  {title} {\bibinfo {title} {Quantum cryptography and
  bell’s theorem},\ }in\ \href@noop {} {\emph {\bibinfo {booktitle} {Quantum
  Measurements in Optics}}}\ (\bibinfo  {publisher} {Springer},\ \bibinfo
  {year} {1992})\ pp.\ \bibinfo {pages} {413--418}\BibitemShut {NoStop}%
\bibitem [{\citenamefont {Scarani}\ \emph {et~al.}(2009)\citenamefont
  {Scarani}, \citenamefont {Bechmann-Pasquinucci}, \citenamefont {Cerf},
  \citenamefont {Du{\v{s}}ek}, \citenamefont {L{\"u}tkenhaus},\ and\
  \citenamefont {Peev}}]{scarani2009security}%
  \BibitemOpen
  \bibfield  {author} {\bibinfo {author} {\bibfnamefont {V.}~\bibnamefont
  {Scarani}}, \bibinfo {author} {\bibfnamefont {H.}~\bibnamefont
  {Bechmann-Pasquinucci}}, \bibinfo {author} {\bibfnamefont {N.~J.}\
  \bibnamefont {Cerf}}, \bibinfo {author} {\bibfnamefont {M.}~\bibnamefont
  {Du{\v{s}}ek}}, \bibinfo {author} {\bibfnamefont {N.}~\bibnamefont
  {L{\"u}tkenhaus}},\ and\ \bibinfo {author} {\bibfnamefont {M.}~\bibnamefont
  {Peev}},\ }\bibfield  {title} {\bibinfo {title} {The security of practical
  quantum key distribution},\ }\href@noop {} {\bibfield  {journal} {\bibinfo
  {journal} {Reviews of modern physics}\ }\textbf {\bibinfo {volume} {81}},\
  \bibinfo {pages} {1301} (\bibinfo {year} {2009})}\BibitemShut {NoStop}%
\bibitem [{\citenamefont {Gharibian}(2008)}]{gharibian2008strong}%
  \BibitemOpen
  \bibfield  {author} {\bibinfo {author} {\bibfnamefont {S.}~\bibnamefont
  {Gharibian}},\ }\bibfield  {title} {\bibinfo {title} {Strong np-hardness of
  the quantum separability problem},\ }\href@noop {} {\bibfield  {journal}
  {\bibinfo  {journal} {arXiv preprint arXiv:0810.4507}\ } (\bibinfo {year}
  {2008})}\BibitemShut {NoStop}%
\bibitem [{\citenamefont {Peres}(1996)}]{peres1996separability}%
  \BibitemOpen
  \bibfield  {author} {\bibinfo {author} {\bibfnamefont {A.}~\bibnamefont
  {Peres}},\ }\bibfield  {title} {\bibinfo {title} {Separability criterion for
  density matrices},\ }\href@noop {} {\bibfield  {journal} {\bibinfo  {journal}
  {Physical Review Letters}\ }\textbf {\bibinfo {volume} {77}},\ \bibinfo
  {pages} {1413} (\bibinfo {year} {1996})}\BibitemShut {NoStop}%
\bibitem [{\citenamefont {Horodecki}(1997)}]{horodecki1997separability}%
  \BibitemOpen
  \bibfield  {author} {\bibinfo {author} {\bibfnamefont {P.}~\bibnamefont
  {Horodecki}},\ }\bibfield  {title} {\bibinfo {title} {Separability criterion
  and inseparable mixed states with positive partial transposition},\
  }\href@noop {} {\bibfield  {journal} {\bibinfo  {journal} {Physics Letters
  A}\ }\textbf {\bibinfo {volume} {232}},\ \bibinfo {pages} {333} (\bibinfo
  {year} {1997})}\BibitemShut {NoStop}%
\bibitem [{\citenamefont {Horodecki}\ \emph {et~al.}(1996)\citenamefont
  {Horodecki}, \citenamefont {Horodecki},\ and\ \citenamefont
  {Horodecki}}]{horodecki1996teleportation}%
  \BibitemOpen
  \bibfield  {author} {\bibinfo {author} {\bibfnamefont {R.}~\bibnamefont
  {Horodecki}}, \bibinfo {author} {\bibfnamefont {M.}~\bibnamefont
  {Horodecki}},\ and\ \bibinfo {author} {\bibfnamefont {P.}~\bibnamefont
  {Horodecki}},\ }\bibfield  {title} {\bibinfo {title} {Teleportation, bell's
  inequalities and inseparability},\ }\href@noop {} {\bibfield  {journal}
  {\bibinfo  {journal} {Physics Letters A}\ }\textbf {\bibinfo {volume}
  {222}},\ \bibinfo {pages} {21} (\bibinfo {year} {1996})}\BibitemShut
  {NoStop}%
\bibitem [{\citenamefont {G{\"u}hne}\ and\ \citenamefont
  {T{\'o}th}(2009)}]{guhne2009entanglement}%
  \BibitemOpen
  \bibfield  {author} {\bibinfo {author} {\bibfnamefont {O.}~\bibnamefont
  {G{\"u}hne}}\ and\ \bibinfo {author} {\bibfnamefont {G.}~\bibnamefont
  {T{\'o}th}},\ }\bibfield  {title} {\bibinfo {title} {Entanglement
  detection},\ }\href@noop {} {\bibfield  {journal} {\bibinfo  {journal}
  {Physics Reports}\ }\textbf {\bibinfo {volume} {474}},\ \bibinfo {pages} {1}
  (\bibinfo {year} {2009})}\BibitemShut {NoStop}%
\bibitem [{\citenamefont {Doherty}\ \emph {et~al.}(2004)\citenamefont
  {Doherty}, \citenamefont {Parrilo},\ and\ \citenamefont
  {Spedalieri}}]{doherty2004complete}%
  \BibitemOpen
  \bibfield  {author} {\bibinfo {author} {\bibfnamefont {A.~C.}\ \bibnamefont
  {Doherty}}, \bibinfo {author} {\bibfnamefont {P.~A.}\ \bibnamefont
  {Parrilo}},\ and\ \bibinfo {author} {\bibfnamefont {F.~M.}\ \bibnamefont
  {Spedalieri}},\ }\bibfield  {title} {\bibinfo {title} {Complete family of
  separability criteria},\ }\href@noop {} {\bibfield  {journal} {\bibinfo
  {journal} {Physical Review A}\ }\textbf {\bibinfo {volume} {69}},\ \bibinfo
  {pages} {022308} (\bibinfo {year} {2004})}\BibitemShut {NoStop}%
\bibitem [{\citenamefont {Paris}\ and\ \citenamefont
  {Rehacek}(2004)}]{paris2004quantum}%
  \BibitemOpen
  \bibfield  {author} {\bibinfo {author} {\bibfnamefont {M.}~\bibnamefont
  {Paris}}\ and\ \bibinfo {author} {\bibfnamefont {J.}~\bibnamefont
  {Rehacek}},\ }\href@noop {} {\emph {\bibinfo {title} {Quantum state
  estimation}}},\ Vol.\ \bibinfo {volume} {649}\ (\bibinfo  {publisher}
  {Springer Science \& Business Media},\ \bibinfo {year} {2004})\BibitemShut
  {NoStop}%
\bibitem [{\citenamefont {Steffen}\ \emph {et~al.}(2006)\citenamefont
  {Steffen}, \citenamefont {Ansmann}, \citenamefont {Bialczak}, \citenamefont
  {Katz}, \citenamefont {Lucero}, \citenamefont {McDermott}, \citenamefont
  {Neeley}, \citenamefont {Weig}, \citenamefont {Cleland},\ and\ \citenamefont
  {Martinis}}]{steffen2006measurement}%
  \BibitemOpen
  \bibfield  {author} {\bibinfo {author} {\bibfnamefont {M.}~\bibnamefont
  {Steffen}}, \bibinfo {author} {\bibfnamefont {M.}~\bibnamefont {Ansmann}},
  \bibinfo {author} {\bibfnamefont {R.~C.}\ \bibnamefont {Bialczak}}, \bibinfo
  {author} {\bibfnamefont {N.}~\bibnamefont {Katz}}, \bibinfo {author}
  {\bibfnamefont {E.}~\bibnamefont {Lucero}}, \bibinfo {author} {\bibfnamefont
  {R.}~\bibnamefont {McDermott}}, \bibinfo {author} {\bibfnamefont
  {M.}~\bibnamefont {Neeley}}, \bibinfo {author} {\bibfnamefont {E.~M.}\
  \bibnamefont {Weig}}, \bibinfo {author} {\bibfnamefont {A.~N.}\ \bibnamefont
  {Cleland}},\ and\ \bibinfo {author} {\bibfnamefont {J.~M.}\ \bibnamefont
  {Martinis}},\ }\bibfield  {title} {\bibinfo {title} {Measurement of the
  entanglement of two superconducting qubits via state tomography},\
  }\href@noop {} {\bibfield  {journal} {\bibinfo  {journal} {Science}\ }\textbf
  {\bibinfo {volume} {313}},\ \bibinfo {pages} {1423} (\bibinfo {year}
  {2006})}\BibitemShut {NoStop}%
\bibitem [{\citenamefont {Poyatos}\ \emph {et~al.}(1997)\citenamefont
  {Poyatos}, \citenamefont {Cirac},\ and\ \citenamefont
  {Zoller}}]{poyatos1997complete}%
  \BibitemOpen
  \bibfield  {author} {\bibinfo {author} {\bibfnamefont {J.}~\bibnamefont
  {Poyatos}}, \bibinfo {author} {\bibfnamefont {J.~I.}\ \bibnamefont {Cirac}},\
  and\ \bibinfo {author} {\bibfnamefont {P.}~\bibnamefont {Zoller}},\
  }\bibfield  {title} {\bibinfo {title} {Complete characterization of a quantum
  process: the two-bit quantum gate},\ }\href@noop {} {\bibfield  {journal}
  {\bibinfo  {journal} {Physical Review Letters}\ }\textbf {\bibinfo {volume}
  {78}},\ \bibinfo {pages} {390} (\bibinfo {year} {1997})}\BibitemShut
  {NoStop}%
\bibitem [{\citenamefont {Bavaresco}\ \emph {et~al.}(2018)\citenamefont
  {Bavaresco}, \citenamefont {Herrera~Valencia}, \citenamefont {Kl{\"o}ckl},
  \citenamefont {Pivoluska}, \citenamefont {Erker}, \citenamefont {Friis},
  \citenamefont {Malik},\ and\ \citenamefont
  {Huber}}]{bavaresco2018measurements}%
  \BibitemOpen
  \bibfield  {author} {\bibinfo {author} {\bibfnamefont {J.}~\bibnamefont
  {Bavaresco}}, \bibinfo {author} {\bibfnamefont {N.}~\bibnamefont
  {Herrera~Valencia}}, \bibinfo {author} {\bibfnamefont {C.}~\bibnamefont
  {Kl{\"o}ckl}}, \bibinfo {author} {\bibfnamefont {M.}~\bibnamefont
  {Pivoluska}}, \bibinfo {author} {\bibfnamefont {P.}~\bibnamefont {Erker}},
  \bibinfo {author} {\bibfnamefont {N.}~\bibnamefont {Friis}}, \bibinfo
  {author} {\bibfnamefont {M.}~\bibnamefont {Malik}},\ and\ \bibinfo {author}
  {\bibfnamefont {M.}~\bibnamefont {Huber}},\ }\bibfield  {title} {\bibinfo
  {title} {Measurements in two bases are sufficient for certifying
  high-dimensional entanglement},\ }\href@noop {} {\bibfield  {journal}
  {\bibinfo  {journal} {Nature Physics}\ }\textbf {\bibinfo {volume} {14}},\
  \bibinfo {pages} {1032} (\bibinfo {year} {2018})}\BibitemShut {NoStop}%
\bibitem [{\citenamefont {Rosset}\ \emph {et~al.}(2012)\citenamefont {Rosset},
  \citenamefont {Ferretti-Sch{\"o}bitz}, \citenamefont {Bancal}, \citenamefont
  {Gisin},\ and\ \citenamefont {Liang}}]{rosset2012imperfect}%
  \BibitemOpen
  \bibfield  {author} {\bibinfo {author} {\bibfnamefont {D.}~\bibnamefont
  {Rosset}}, \bibinfo {author} {\bibfnamefont {R.}~\bibnamefont
  {Ferretti-Sch{\"o}bitz}}, \bibinfo {author} {\bibfnamefont {J.-D.}\
  \bibnamefont {Bancal}}, \bibinfo {author} {\bibfnamefont {N.}~\bibnamefont
  {Gisin}},\ and\ \bibinfo {author} {\bibfnamefont {Y.-C.}\ \bibnamefont
  {Liang}},\ }\bibfield  {title} {\bibinfo {title} {Imperfect measurement
  settings: Implications for quantum state tomography and entanglement
  witnesses},\ }\href@noop {} {\bibfield  {journal} {\bibinfo  {journal}
  {Physical Review A}\ }\textbf {\bibinfo {volume} {86}},\ \bibinfo {pages}
  {062325} (\bibinfo {year} {2012})}\BibitemShut {NoStop}%
\bibitem [{\citenamefont {Friis}\ \emph {et~al.}(2019)\citenamefont {Friis},
  \citenamefont {Vitagliano}, \citenamefont {Malik},\ and\ \citenamefont
  {Huber}}]{friis2019entanglement}%
  \BibitemOpen
  \bibfield  {author} {\bibinfo {author} {\bibfnamefont {N.}~\bibnamefont
  {Friis}}, \bibinfo {author} {\bibfnamefont {G.}~\bibnamefont {Vitagliano}},
  \bibinfo {author} {\bibfnamefont {M.}~\bibnamefont {Malik}},\ and\ \bibinfo
  {author} {\bibfnamefont {M.}~\bibnamefont {Huber}},\ }\bibfield  {title}
  {\bibinfo {title} {Entanglement certification from theory to experiment},\
  }\href@noop {} {\bibfield  {journal} {\bibinfo  {journal} {Nature Reviews
  Physics}\ }\textbf {\bibinfo {volume} {1}},\ \bibinfo {pages} {72} (\bibinfo
  {year} {2019})}\BibitemShut {NoStop}%
\bibitem [{\citenamefont {Collins}\ \emph {et~al.}(2002)\citenamefont
  {Collins}, \citenamefont {Gisin}, \citenamefont {Linden}, \citenamefont
  {Massar},\ and\ \citenamefont {Popescu}}]{collins2002bell}%
  \BibitemOpen
  \bibfield  {author} {\bibinfo {author} {\bibfnamefont {D.}~\bibnamefont
  {Collins}}, \bibinfo {author} {\bibfnamefont {N.}~\bibnamefont {Gisin}},
  \bibinfo {author} {\bibfnamefont {N.}~\bibnamefont {Linden}}, \bibinfo
  {author} {\bibfnamefont {S.}~\bibnamefont {Massar}},\ and\ \bibinfo {author}
  {\bibfnamefont {S.}~\bibnamefont {Popescu}},\ }\bibfield  {title} {\bibinfo
  {title} {Bell inequalities for arbitrarily high-dimensional systems},\
  }\href@noop {} {\bibfield  {journal} {\bibinfo  {journal} {Physical review
  letters}\ }\textbf {\bibinfo {volume} {88}},\ \bibinfo {pages} {040404}
  (\bibinfo {year} {2002})}\BibitemShut {NoStop}%
\bibitem [{\citenamefont {Moroder}\ \emph {et~al.}(2013)\citenamefont
  {Moroder}, \citenamefont {Bancal}, \citenamefont {Liang}, \citenamefont
  {Hofmann},\ and\ \citenamefont {G{\"u}hne}}]{moroder2013device}%
  \BibitemOpen
  \bibfield  {author} {\bibinfo {author} {\bibfnamefont {T.}~\bibnamefont
  {Moroder}}, \bibinfo {author} {\bibfnamefont {J.-D.}\ \bibnamefont {Bancal}},
  \bibinfo {author} {\bibfnamefont {Y.-C.}\ \bibnamefont {Liang}}, \bibinfo
  {author} {\bibfnamefont {M.}~\bibnamefont {Hofmann}},\ and\ \bibinfo {author}
  {\bibfnamefont {O.}~\bibnamefont {G{\"u}hne}},\ }\bibfield  {title} {\bibinfo
  {title} {Device-independent entanglement quantification and related
  applications},\ }\href@noop {} {\bibfield  {journal} {\bibinfo  {journal}
  {Physical review letters}\ }\textbf {\bibinfo {volume} {111}},\ \bibinfo
  {pages} {030501} (\bibinfo {year} {2013})}\BibitemShut {NoStop}%
\bibitem [{\citenamefont {Bowles}\ \emph {et~al.}(2018)\citenamefont {Bowles},
  \citenamefont {{\v{S}}upi{\'c}}, \citenamefont {Cavalcanti},\ and\
  \citenamefont {Ac{\'\i}n}}]{bowles2018device}%
  \BibitemOpen
  \bibfield  {author} {\bibinfo {author} {\bibfnamefont {J.}~\bibnamefont
  {Bowles}}, \bibinfo {author} {\bibfnamefont {I.}~\bibnamefont
  {{\v{S}}upi{\'c}}}, \bibinfo {author} {\bibfnamefont {D.}~\bibnamefont
  {Cavalcanti}},\ and\ \bibinfo {author} {\bibfnamefont {A.}~\bibnamefont
  {Ac{\'\i}n}},\ }\bibfield  {title} {\bibinfo {title} {Device-independent
  entanglement certification of all entangled states},\ }\href@noop {}
  {\bibfield  {journal} {\bibinfo  {journal} {Physical review letters}\
  }\textbf {\bibinfo {volume} {121}},\ \bibinfo {pages} {180503} (\bibinfo
  {year} {2018})}\BibitemShut {NoStop}%
\bibitem [{\citenamefont {Xu}\ \emph {et~al.}(2014)\citenamefont {Xu},
  \citenamefont {Yuan}, \citenamefont {Chen}, \citenamefont {Lu}, \citenamefont
  {Yao}, \citenamefont {Ma}, \citenamefont {Chen},\ and\ \citenamefont
  {Pan}}]{xu2014implementation}%
  \BibitemOpen
  \bibfield  {author} {\bibinfo {author} {\bibfnamefont {P.}~\bibnamefont
  {Xu}}, \bibinfo {author} {\bibfnamefont {X.}~\bibnamefont {Yuan}}, \bibinfo
  {author} {\bibfnamefont {L.-K.}\ \bibnamefont {Chen}}, \bibinfo {author}
  {\bibfnamefont {H.}~\bibnamefont {Lu}}, \bibinfo {author} {\bibfnamefont
  {X.-C.}\ \bibnamefont {Yao}}, \bibinfo {author} {\bibfnamefont
  {X.}~\bibnamefont {Ma}}, \bibinfo {author} {\bibfnamefont {Y.-A.}\
  \bibnamefont {Chen}},\ and\ \bibinfo {author} {\bibfnamefont {J.-W.}\
  \bibnamefont {Pan}},\ }\bibfield  {title} {\bibinfo {title} {Implementation
  of a measurement-device-independent entanglement witness},\ }\href@noop {}
  {\bibfield  {journal} {\bibinfo  {journal} {Physical review letters}\
  }\textbf {\bibinfo {volume} {112}},\ \bibinfo {pages} {140506} (\bibinfo
  {year} {2014})}\BibitemShut {NoStop}%
\bibitem [{\citenamefont {Yeo}\ and\ \citenamefont
  {Chua}(2006)}]{yeo2006teleportation}%
  \BibitemOpen
  \bibfield  {author} {\bibinfo {author} {\bibfnamefont {Y.}~\bibnamefont
  {Yeo}}\ and\ \bibinfo {author} {\bibfnamefont {W.~K.}\ \bibnamefont {Chua}},\
  }\bibfield  {title} {\bibinfo {title} {Teleportation and dense coding with
  genuine multipartite entanglement},\ }\href@noop {} {\bibfield  {journal}
  {\bibinfo  {journal} {Physical Review Letters}\ }\textbf {\bibinfo {volume}
  {96}},\ \bibinfo {pages} {060502} (\bibinfo {year} {2006})}\BibitemShut
  {NoStop}%
\bibitem [{\citenamefont {Chen}\ \emph {et~al.}(2006)\citenamefont {Chen},
  \citenamefont {Zhu},\ and\ \citenamefont {Guo}}]{chen2006general}%
  \BibitemOpen
  \bibfield  {author} {\bibinfo {author} {\bibfnamefont {P.-X.}\ \bibnamefont
  {Chen}}, \bibinfo {author} {\bibfnamefont {S.-Y.}\ \bibnamefont {Zhu}},\ and\
  \bibinfo {author} {\bibfnamefont {G.-C.}\ \bibnamefont {Guo}},\ }\bibfield
  {title} {\bibinfo {title} {General form of genuine multipartite entanglement
  quantum channels for teleportation},\ }\href@noop {} {\bibfield  {journal}
  {\bibinfo  {journal} {Physical Review A}\ }\textbf {\bibinfo {volume} {74}},\
  \bibinfo {pages} {032324} (\bibinfo {year} {2006})}\BibitemShut {NoStop}%
\bibitem [{\citenamefont {Muralidharan}\ and\ \citenamefont
  {Panigrahi}(2008)}]{muralidharan2008perfect}%
  \BibitemOpen
  \bibfield  {author} {\bibinfo {author} {\bibfnamefont {S.}~\bibnamefont
  {Muralidharan}}\ and\ \bibinfo {author} {\bibfnamefont {P.~K.}\ \bibnamefont
  {Panigrahi}},\ }\bibfield  {title} {\bibinfo {title} {Perfect teleportation,
  quantum-state sharing, and superdense coding through a genuinely entangled
  five-qubit state},\ }\href@noop {} {\bibfield  {journal} {\bibinfo  {journal}
  {Physical Review A}\ }\textbf {\bibinfo {volume} {77}},\ \bibinfo {pages}
  {032321} (\bibinfo {year} {2008})}\BibitemShut {NoStop}%
\bibitem [{\citenamefont {T{\'o}th}(2012)}]{toth2012multipartite}%
  \BibitemOpen
  \bibfield  {author} {\bibinfo {author} {\bibfnamefont {G.}~\bibnamefont
  {T{\'o}th}},\ }\bibfield  {title} {\bibinfo {title} {Multipartite
  entanglement and high-precision metrology},\ }\href@noop {} {\bibfield
  {journal} {\bibinfo  {journal} {Physical Review A}\ }\textbf {\bibinfo
  {volume} {85}},\ \bibinfo {pages} {022322} (\bibinfo {year}
  {2012})}\BibitemShut {NoStop}%
\bibitem [{\citenamefont {Hyllus}\ \emph {et~al.}(2012)\citenamefont {Hyllus},
  \citenamefont {Laskowski}, \citenamefont {Krischek}, \citenamefont
  {Schwemmer}, \citenamefont {Wieczorek}, \citenamefont {Weinfurter},
  \citenamefont {Pezz{\'e}},\ and\ \citenamefont {Smerzi}}]{hyllus2012fisher}%
  \BibitemOpen
  \bibfield  {author} {\bibinfo {author} {\bibfnamefont {P.}~\bibnamefont
  {Hyllus}}, \bibinfo {author} {\bibfnamefont {W.}~\bibnamefont {Laskowski}},
  \bibinfo {author} {\bibfnamefont {R.}~\bibnamefont {Krischek}}, \bibinfo
  {author} {\bibfnamefont {C.}~\bibnamefont {Schwemmer}}, \bibinfo {author}
  {\bibfnamefont {W.}~\bibnamefont {Wieczorek}}, \bibinfo {author}
  {\bibfnamefont {H.}~\bibnamefont {Weinfurter}}, \bibinfo {author}
  {\bibfnamefont {L.}~\bibnamefont {Pezz{\'e}}},\ and\ \bibinfo {author}
  {\bibfnamefont {A.}~\bibnamefont {Smerzi}},\ }\bibfield  {title} {\bibinfo
  {title} {Fisher information and multiparticle entanglement},\ }\href@noop {}
  {\bibfield  {journal} {\bibinfo  {journal} {Physical Review A}\ }\textbf
  {\bibinfo {volume} {85}},\ \bibinfo {pages} {022321} (\bibinfo {year}
  {2012})}\BibitemShut {NoStop}%
\bibitem [{\citenamefont {Sarovar}\ \emph {et~al.}(2010)\citenamefont
  {Sarovar}, \citenamefont {Ishizaki}, \citenamefont {Fleming},\ and\
  \citenamefont {Whaley}}]{sarovar2010quantum}%
  \BibitemOpen
  \bibfield  {author} {\bibinfo {author} {\bibfnamefont {M.}~\bibnamefont
  {Sarovar}}, \bibinfo {author} {\bibfnamefont {A.}~\bibnamefont {Ishizaki}},
  \bibinfo {author} {\bibfnamefont {G.~R.}\ \bibnamefont {Fleming}},\ and\
  \bibinfo {author} {\bibfnamefont {K.~B.}\ \bibnamefont {Whaley}},\ }\bibfield
   {title} {\bibinfo {title} {Quantum entanglement in photosynthetic
  light-harvesting complexes},\ }\href@noop {} {\bibfield  {journal} {\bibinfo
  {journal} {Nature Physics}\ }\textbf {\bibinfo {volume} {6}},\ \bibinfo
  {pages} {462} (\bibinfo {year} {2010})}\BibitemShut {NoStop}%
\bibitem [{\citenamefont {G{\"u}hne}\ and\ \citenamefont
  {Seevinck}(2010)}]{guhne2010separability}%
  \BibitemOpen
  \bibfield  {author} {\bibinfo {author} {\bibfnamefont {O.}~\bibnamefont
  {G{\"u}hne}}\ and\ \bibinfo {author} {\bibfnamefont {M.}~\bibnamefont
  {Seevinck}},\ }\bibfield  {title} {\bibinfo {title} {Separability criteria
  for genuine multiparticle entanglement},\ }\href@noop {} {\bibfield
  {journal} {\bibinfo  {journal} {New Journal of Physics}\ }\textbf {\bibinfo
  {volume} {12}},\ \bibinfo {pages} {053002} (\bibinfo {year}
  {2010})}\BibitemShut {NoStop}%
\bibitem [{\citenamefont {Szalay}(2015)}]{szalay2015multipartite}%
  \BibitemOpen
  \bibfield  {author} {\bibinfo {author} {\bibfnamefont {S.}~\bibnamefont
  {Szalay}},\ }\bibfield  {title} {\bibinfo {title} {Multipartite entanglement
  measures},\ }\href@noop {} {\bibfield  {journal} {\bibinfo  {journal}
  {Physical Review A}\ }\textbf {\bibinfo {volume} {92}},\ \bibinfo {pages}
  {042329} (\bibinfo {year} {2015})}\BibitemShut {NoStop}%
\bibitem [{\citenamefont {Ma}\ \emph {et~al.}(2011)\citenamefont {Ma},
  \citenamefont {Chen}, \citenamefont {Chen}, \citenamefont {Spengler},
  \citenamefont {Gabriel},\ and\ \citenamefont {Huber}}]{ma2011measure}%
  \BibitemOpen
  \bibfield  {author} {\bibinfo {author} {\bibfnamefont {Z.-H.}\ \bibnamefont
  {Ma}}, \bibinfo {author} {\bibfnamefont {Z.-H.}\ \bibnamefont {Chen}},
  \bibinfo {author} {\bibfnamefont {J.-L.}\ \bibnamefont {Chen}}, \bibinfo
  {author} {\bibfnamefont {C.}~\bibnamefont {Spengler}}, \bibinfo {author}
  {\bibfnamefont {A.}~\bibnamefont {Gabriel}},\ and\ \bibinfo {author}
  {\bibfnamefont {M.}~\bibnamefont {Huber}},\ }\bibfield  {title} {\bibinfo
  {title} {Measure of genuine multipartite entanglement with computable lower
  bounds},\ }\href@noop {} {\bibfield  {journal} {\bibinfo  {journal} {Physical
  Review A}\ }\textbf {\bibinfo {volume} {83}},\ \bibinfo {pages} {062325}
  (\bibinfo {year} {2011})}\BibitemShut {NoStop}%
\bibitem [{\citenamefont {Chen}\ \emph {et~al.}(2012)\citenamefont {Chen},
  \citenamefont {Ma}, \citenamefont {Chen},\ and\ \citenamefont
  {Severini}}]{chen2012improved}%
  \BibitemOpen
  \bibfield  {author} {\bibinfo {author} {\bibfnamefont {Z.-H.}\ \bibnamefont
  {Chen}}, \bibinfo {author} {\bibfnamefont {Z.-H.}\ \bibnamefont {Ma}},
  \bibinfo {author} {\bibfnamefont {J.-L.}\ \bibnamefont {Chen}},\ and\
  \bibinfo {author} {\bibfnamefont {S.}~\bibnamefont {Severini}},\ }\bibfield
  {title} {\bibinfo {title} {Improved lower bounds on
  genuine-multipartite-entanglement concurrence},\ }\href@noop {} {\bibfield
  {journal} {\bibinfo  {journal} {Physical Review A}\ }\textbf {\bibinfo
  {volume} {85}},\ \bibinfo {pages} {062320} (\bibinfo {year}
  {2012})}\BibitemShut {NoStop}%
\bibitem [{\citenamefont {de~Vicente}\ and\ \citenamefont
  {Huber}(2011)}]{de2011multipartite}%
  \BibitemOpen
  \bibfield  {author} {\bibinfo {author} {\bibfnamefont {J.~I.}\ \bibnamefont
  {de~Vicente}}\ and\ \bibinfo {author} {\bibfnamefont {M.}~\bibnamefont
  {Huber}},\ }\bibfield  {title} {\bibinfo {title} {Multipartite entanglement
  detection from correlation tensors},\ }\href@noop {} {\bibfield  {journal}
  {\bibinfo  {journal} {Physical Review A}\ }\textbf {\bibinfo {volume} {84}},\
  \bibinfo {pages} {062306} (\bibinfo {year} {2011})}\BibitemShut {NoStop}%
\bibitem [{\citenamefont {Huber}\ \emph {et~al.}(2010)\citenamefont {Huber},
  \citenamefont {Mintert}, \citenamefont {Gabriel},\ and\ \citenamefont
  {Hiesmayr}}]{huber2010detection}%
  \BibitemOpen
  \bibfield  {author} {\bibinfo {author} {\bibfnamefont {M.}~\bibnamefont
  {Huber}}, \bibinfo {author} {\bibfnamefont {F.}~\bibnamefont {Mintert}},
  \bibinfo {author} {\bibfnamefont {A.}~\bibnamefont {Gabriel}},\ and\ \bibinfo
  {author} {\bibfnamefont {B.~C.}\ \bibnamefont {Hiesmayr}},\ }\bibfield
  {title} {\bibinfo {title} {Detection of high-dimensional genuine multipartite
  entanglement of mixed states},\ }\href@noop {} {\bibfield  {journal}
  {\bibinfo  {journal} {Physical review letters}\ }\textbf {\bibinfo {volume}
  {104}},\ \bibinfo {pages} {210501} (\bibinfo {year} {2010})}\BibitemShut
  {NoStop}%
\bibitem [{\citenamefont {Jungnitsch}\ \emph {et~al.}(2011)\citenamefont
  {Jungnitsch}, \citenamefont {Moroder},\ and\ \citenamefont
  {G{\"u}hne}}]{jungnitsch2011taming}%
  \BibitemOpen
  \bibfield  {author} {\bibinfo {author} {\bibfnamefont {B.}~\bibnamefont
  {Jungnitsch}}, \bibinfo {author} {\bibfnamefont {T.}~\bibnamefont
  {Moroder}},\ and\ \bibinfo {author} {\bibfnamefont {O.}~\bibnamefont
  {G{\"u}hne}},\ }\bibfield  {title} {\bibinfo {title} {Taming multiparticle
  entanglement},\ }\href@noop {} {\bibfield  {journal} {\bibinfo  {journal}
  {Physical review letters}\ }\textbf {\bibinfo {volume} {106}},\ \bibinfo
  {pages} {190502} (\bibinfo {year} {2011})}\BibitemShut {NoStop}%
\bibitem [{\citenamefont {Bancal}\ \emph {et~al.}(2011)\citenamefont {Bancal},
  \citenamefont {Gisin}, \citenamefont {Liang},\ and\ \citenamefont
  {Pironio}}]{bancal2011device}%
  \BibitemOpen
  \bibfield  {author} {\bibinfo {author} {\bibfnamefont {J.-D.}\ \bibnamefont
  {Bancal}}, \bibinfo {author} {\bibfnamefont {N.}~\bibnamefont {Gisin}},
  \bibinfo {author} {\bibfnamefont {Y.-C.}\ \bibnamefont {Liang}},\ and\
  \bibinfo {author} {\bibfnamefont {S.}~\bibnamefont {Pironio}},\ }\bibfield
  {title} {\bibinfo {title} {Device-independent witnesses of genuine
  multipartite entanglement},\ }\href@noop {} {\bibfield  {journal} {\bibinfo
  {journal} {Physical Review Letters}\ }\textbf {\bibinfo {volume} {106}},\
  \bibinfo {pages} {250404} (\bibinfo {year} {2011})}\BibitemShut {NoStop}%
\bibitem [{\citenamefont {T{\'o}th}\ and\ \citenamefont
  {G{\"u}hne}(2005)}]{toth2005detecting}%
  \BibitemOpen
  \bibfield  {author} {\bibinfo {author} {\bibfnamefont {G.}~\bibnamefont
  {T{\'o}th}}\ and\ \bibinfo {author} {\bibfnamefont {O.}~\bibnamefont
  {G{\"u}hne}},\ }\bibfield  {title} {\bibinfo {title} {Detecting genuine
  multipartite entanglement with two local measurements},\ }\href@noop {}
  {\bibfield  {journal} {\bibinfo  {journal} {Physical review letters}\
  }\textbf {\bibinfo {volume} {94}},\ \bibinfo {pages} {060501} (\bibinfo
  {year} {2005})}\BibitemShut {NoStop}%
\bibitem [{\citenamefont {Bourennane}\ \emph {et~al.}(2004)\citenamefont
  {Bourennane}, \citenamefont {Eibl}, \citenamefont {Kurtsiefer}, \citenamefont
  {Gaertner}, \citenamefont {Weinfurter}, \citenamefont {G{\"u}hne},
  \citenamefont {Hyllus}, \citenamefont {Bru{\ss}}, \citenamefont
  {Lewenstein},\ and\ \citenamefont {Sanpera}}]{bourennane2004experimental}%
  \BibitemOpen
  \bibfield  {author} {\bibinfo {author} {\bibfnamefont {M.}~\bibnamefont
  {Bourennane}}, \bibinfo {author} {\bibfnamefont {M.}~\bibnamefont {Eibl}},
  \bibinfo {author} {\bibfnamefont {C.}~\bibnamefont {Kurtsiefer}}, \bibinfo
  {author} {\bibfnamefont {S.}~\bibnamefont {Gaertner}}, \bibinfo {author}
  {\bibfnamefont {H.}~\bibnamefont {Weinfurter}}, \bibinfo {author}
  {\bibfnamefont {O.}~\bibnamefont {G{\"u}hne}}, \bibinfo {author}
  {\bibfnamefont {P.}~\bibnamefont {Hyllus}}, \bibinfo {author} {\bibfnamefont
  {D.}~\bibnamefont {Bru{\ss}}}, \bibinfo {author} {\bibfnamefont
  {M.}~\bibnamefont {Lewenstein}},\ and\ \bibinfo {author} {\bibfnamefont
  {A.}~\bibnamefont {Sanpera}},\ }\bibfield  {title} {\bibinfo {title}
  {Experimental detection of multipartite entanglement using witness
  operators},\ }\href@noop {} {\bibfield  {journal} {\bibinfo  {journal}
  {Physical review letters}\ }\textbf {\bibinfo {volume} {92}},\ \bibinfo
  {pages} {087902} (\bibinfo {year} {2004})}\BibitemShut {NoStop}%
\bibitem [{\citenamefont {P{\'a}l}\ and\ \citenamefont
  {V{\'e}rtesi}(2011)}]{pal2011multisetting}%
  \BibitemOpen
  \bibfield  {author} {\bibinfo {author} {\bibfnamefont {K.~F.}\ \bibnamefont
  {P{\'a}l}}\ and\ \bibinfo {author} {\bibfnamefont {T.}~\bibnamefont
  {V{\'e}rtesi}},\ }\bibfield  {title} {\bibinfo {title} {Multisetting
  bell-type inequalities for detecting genuine multipartite entanglement},\
  }\href@noop {} {\bibfield  {journal} {\bibinfo  {journal} {Physical Review
  A}\ }\textbf {\bibinfo {volume} {83}},\ \bibinfo {pages} {062123} (\bibinfo
  {year} {2011})}\BibitemShut {NoStop}%
\bibitem [{\citenamefont {Barreiro}\ \emph {et~al.}(2013)\citenamefont
  {Barreiro}, \citenamefont {Bancal}, \citenamefont {Schindler}, \citenamefont
  {Nigg}, \citenamefont {Hennrich}, \citenamefont {Monz}, \citenamefont
  {Gisin},\ and\ \citenamefont {Blatt}}]{barreiro2013demonstration}%
  \BibitemOpen
  \bibfield  {author} {\bibinfo {author} {\bibfnamefont {J.~T.}\ \bibnamefont
  {Barreiro}}, \bibinfo {author} {\bibfnamefont {J.-D.}\ \bibnamefont
  {Bancal}}, \bibinfo {author} {\bibfnamefont {P.}~\bibnamefont {Schindler}},
  \bibinfo {author} {\bibfnamefont {D.}~\bibnamefont {Nigg}}, \bibinfo {author}
  {\bibfnamefont {M.}~\bibnamefont {Hennrich}}, \bibinfo {author}
  {\bibfnamefont {T.}~\bibnamefont {Monz}}, \bibinfo {author} {\bibfnamefont
  {N.}~\bibnamefont {Gisin}},\ and\ \bibinfo {author} {\bibfnamefont
  {R.}~\bibnamefont {Blatt}},\ }\bibfield  {title} {\bibinfo {title}
  {Demonstration of genuine multipartite entanglement with device-independent
  witnesses},\ }\href@noop {} {\bibfield  {journal} {\bibinfo  {journal}
  {Nature Physics}\ }\textbf {\bibinfo {volume} {9}},\ \bibinfo {pages} {559}
  (\bibinfo {year} {2013})}\BibitemShut {NoStop}%
\bibitem [{\citenamefont {Girardin}\ \emph {et~al.}(2022)\citenamefont
  {Girardin}, \citenamefont {Brunner},\ and\ \citenamefont
  {Kriv{\'a}chy}}]{girardin2022building}%
  \BibitemOpen
  \bibfield  {author} {\bibinfo {author} {\bibfnamefont {A.}~\bibnamefont
  {Girardin}}, \bibinfo {author} {\bibfnamefont {N.}~\bibnamefont {Brunner}},\
  and\ \bibinfo {author} {\bibfnamefont {T.}~\bibnamefont {Kriv{\'a}chy}},\
  }\bibfield  {title} {\bibinfo {title} {Building separable approximations for
  quantum states via neural networks},\ }\href@noop {} {\bibfield  {journal}
  {\bibinfo  {journal} {Physical Review Research}\ }\textbf {\bibinfo {volume}
  {4}},\ \bibinfo {pages} {023238} (\bibinfo {year} {2022})}\BibitemShut
  {NoStop}%
\bibitem [{\citenamefont {Lu}\ \emph {et~al.}(2018{\natexlab{a}})\citenamefont
  {Lu}, \citenamefont {Huang}, \citenamefont {Li}, \citenamefont {Li},
  \citenamefont {Chen}, \citenamefont {Lu}, \citenamefont {Ji}, \citenamefont
  {Shen}, \citenamefont {Zhou},\ and\ \citenamefont
  {Zeng}}]{lu2018separability}%
  \BibitemOpen
  \bibfield  {author} {\bibinfo {author} {\bibfnamefont {S.}~\bibnamefont
  {Lu}}, \bibinfo {author} {\bibfnamefont {S.}~\bibnamefont {Huang}}, \bibinfo
  {author} {\bibfnamefont {K.}~\bibnamefont {Li}}, \bibinfo {author}
  {\bibfnamefont {J.}~\bibnamefont {Li}}, \bibinfo {author} {\bibfnamefont
  {J.}~\bibnamefont {Chen}}, \bibinfo {author} {\bibfnamefont {D.}~\bibnamefont
  {Lu}}, \bibinfo {author} {\bibfnamefont {Z.}~\bibnamefont {Ji}}, \bibinfo
  {author} {\bibfnamefont {Y.}~\bibnamefont {Shen}}, \bibinfo {author}
  {\bibfnamefont {D.}~\bibnamefont {Zhou}},\ and\ \bibinfo {author}
  {\bibfnamefont {B.}~\bibnamefont {Zeng}},\ }\bibfield  {title} {\bibinfo
  {title} {Separability-entanglement classifier via machine learning},\
  }\href@noop {} {\bibfield  {journal} {\bibinfo  {journal} {Physical Review
  A}\ }\textbf {\bibinfo {volume} {98}},\ \bibinfo {pages} {012315} (\bibinfo
  {year} {2018}{\natexlab{a}})}\BibitemShut {NoStop}%
\bibitem [{\citenamefont {Chen}\ \emph
  {et~al.}(2021{\natexlab{a}})\citenamefont {Chen}, \citenamefont {Pan},
  \citenamefont {Zhang},\ and\ \citenamefont {Cheng}}]{chen2021detecting}%
  \BibitemOpen
  \bibfield  {author} {\bibinfo {author} {\bibfnamefont {Y.}~\bibnamefont
  {Chen}}, \bibinfo {author} {\bibfnamefont {Y.}~\bibnamefont {Pan}}, \bibinfo
  {author} {\bibfnamefont {G.}~\bibnamefont {Zhang}},\ and\ \bibinfo {author}
  {\bibfnamefont {S.}~\bibnamefont {Cheng}},\ }\bibfield  {title} {\bibinfo
  {title} {Detecting quantum entanglement with unsupervised learning},\
  }\href@noop {} {\bibfield  {journal} {\bibinfo  {journal} {Quantum Science
  and Technology}\ }\textbf {\bibinfo {volume} {7}},\ \bibinfo {pages} {015005}
  (\bibinfo {year} {2021}{\natexlab{a}})}\BibitemShut {NoStop}%
\bibitem [{\citenamefont {Chen}\ \emph
  {et~al.}(2021{\natexlab{b}})\citenamefont {Chen}, \citenamefont {Ren},
  \citenamefont {Lin},\ and\ \citenamefont {Lu}}]{chen2021entanglement}%
  \BibitemOpen
  \bibfield  {author} {\bibinfo {author} {\bibfnamefont {C.}~\bibnamefont
  {Chen}}, \bibinfo {author} {\bibfnamefont {C.}~\bibnamefont {Ren}}, \bibinfo
  {author} {\bibfnamefont {H.}~\bibnamefont {Lin}},\ and\ \bibinfo {author}
  {\bibfnamefont {H.}~\bibnamefont {Lu}},\ }\bibfield  {title} {\bibinfo
  {title} {Entanglement structure detection via machine learning},\ }\href@noop
  {} {\bibfield  {journal} {\bibinfo  {journal} {Quantum Science and
  Technology}\ } (\bibinfo {year} {2021}{\natexlab{b}})}\BibitemShut {NoStop}%
\bibitem [{\citenamefont {Ren}\ and\ \citenamefont
  {Chen}(2019)}]{ren2019steerability}%
  \BibitemOpen
  \bibfield  {author} {\bibinfo {author} {\bibfnamefont {C.}~\bibnamefont
  {Ren}}\ and\ \bibinfo {author} {\bibfnamefont {C.}~\bibnamefont {Chen}},\
  }\bibfield  {title} {\bibinfo {title} {Steerability detection of an arbitrary
  two-qubit state via machine learning},\ }\href@noop {} {\bibfield  {journal}
  {\bibinfo  {journal} {Physical Review A}\ }\textbf {\bibinfo {volume}
  {100}},\ \bibinfo {pages} {022314} (\bibinfo {year} {2019})}\BibitemShut
  {NoStop}%
\bibitem [{\citenamefont {Canabarro}\ \emph {et~al.}(2019)\citenamefont
  {Canabarro}, \citenamefont {Brito},\ and\ \citenamefont
  {Chaves}}]{canabarro2019machine}%
  \BibitemOpen
  \bibfield  {author} {\bibinfo {author} {\bibfnamefont {A.}~\bibnamefont
  {Canabarro}}, \bibinfo {author} {\bibfnamefont {S.}~\bibnamefont {Brito}},\
  and\ \bibinfo {author} {\bibfnamefont {R.}~\bibnamefont {Chaves}},\
  }\bibfield  {title} {\bibinfo {title} {Machine learning nonlocal
  correlations},\ }\href@noop {} {\bibfield  {journal} {\bibinfo  {journal}
  {Physical review letters}\ }\textbf {\bibinfo {volume} {122}},\ \bibinfo
  {pages} {200401} (\bibinfo {year} {2019})}\BibitemShut {NoStop}%
\bibitem [{\citenamefont {Lin}\ \emph {et~al.}(2021)\citenamefont {Lin},
  \citenamefont {Chen},\ and\ \citenamefont {Wei}}]{lin2021quantifying}%
  \BibitemOpen
  \bibfield  {author} {\bibinfo {author} {\bibfnamefont {X.}~\bibnamefont
  {Lin}}, \bibinfo {author} {\bibfnamefont {Z.}~\bibnamefont {Chen}},\ and\
  \bibinfo {author} {\bibfnamefont {Z.}~\bibnamefont {Wei}},\ }\bibfield
  {title} {\bibinfo {title} {Quantifying unknown entanglement by neural
  networks},\ }\href@noop {} {\bibfield  {journal} {\bibinfo  {journal} {arXiv
  preprint arXiv:2104.12527}\ } (\bibinfo {year} {2021})}\BibitemShut {NoStop}%
\bibitem [{\citenamefont {Lin}\ \emph {et~al.}(2022)\citenamefont {Lin},
  \citenamefont {Chen},\ and\ \citenamefont {Wei}}]{lin2022quantifying}%
  \BibitemOpen
  \bibfield  {author} {\bibinfo {author} {\bibfnamefont {X.}~\bibnamefont
  {Lin}}, \bibinfo {author} {\bibfnamefont {Z.}~\bibnamefont {Chen}},\ and\
  \bibinfo {author} {\bibfnamefont {Z.}~\bibnamefont {Wei}},\ }\bibfield
  {title} {\bibinfo {title} {Quantifying unknown quantum entanglement via a
  hybrid quantum-classical machine learning framework},\ }\href@noop {}
  {\bibfield  {journal} {\bibinfo  {journal} {arXiv preprint arXiv:2204.11500}\
  } (\bibinfo {year} {2022})}\BibitemShut {NoStop}%
\bibitem [{\citenamefont {CireAan}\ \emph {et~al.}(2012)\citenamefont
  {CireAan}, \citenamefont {Meier}, \citenamefont {Masci},\ and\ \citenamefont
  {Schmidhuber}}]{cireaan2012multi}%
  \BibitemOpen
  \bibfield  {author} {\bibinfo {author} {\bibfnamefont {D.}~\bibnamefont
  {CireAan}}, \bibinfo {author} {\bibfnamefont {U.}~\bibnamefont {Meier}},
  \bibinfo {author} {\bibfnamefont {J.}~\bibnamefont {Masci}},\ and\ \bibinfo
  {author} {\bibfnamefont {J.}~\bibnamefont {Schmidhuber}},\ }\bibfield
  {title} {\bibinfo {title} {Multi-column deep neural network for traffic sign
  classification},\ }\href@noop {} {\bibfield  {journal} {\bibinfo  {journal}
  {Neural networks}\ }\textbf {\bibinfo {volume} {32}},\ \bibinfo {pages} {333}
  (\bibinfo {year} {2012})}\BibitemShut {NoStop}%
\bibitem [{\citenamefont {Gers}\ and\ \citenamefont
  {Schmidhuber}(2001)}]{gers2001lstm}%
  \BibitemOpen
  \bibfield  {author} {\bibinfo {author} {\bibfnamefont {F.~A.}\ \bibnamefont
  {Gers}}\ and\ \bibinfo {author} {\bibfnamefont {E.}~\bibnamefont
  {Schmidhuber}},\ }\bibfield  {title} {\bibinfo {title} {Lstm recurrent
  networks learn simple context-free and context-sensitive languages},\
  }\href@noop {} {\bibfield  {journal} {\bibinfo  {journal} {IEEE Transactions
  on Neural Networks}\ }\textbf {\bibinfo {volume} {12}},\ \bibinfo {pages}
  {1333} (\bibinfo {year} {2001})}\BibitemShut {NoStop}%
\bibitem [{\citenamefont {Elkahky}\ \emph {et~al.}(2015)\citenamefont
  {Elkahky}, \citenamefont {Song},\ and\ \citenamefont
  {He}}]{elkahky2015multi}%
  \BibitemOpen
  \bibfield  {author} {\bibinfo {author} {\bibfnamefont {A.~M.}\ \bibnamefont
  {Elkahky}}, \bibinfo {author} {\bibfnamefont {Y.}~\bibnamefont {Song}},\ and\
  \bibinfo {author} {\bibfnamefont {X.}~\bibnamefont {He}},\ }\bibfield
  {title} {\bibinfo {title} {A multi-view deep learning approach for cross
  domain user modeling in recommendation systems},\ }in\ \href@noop {} {\emph
  {\bibinfo {booktitle} {Proceedings of the 24th international conference on
  world wide web}}}\ (\bibinfo {year} {2015})\ pp.\ \bibinfo {pages}
  {278--288}\BibitemShut {NoStop}%
\bibitem [{\citenamefont {LeCun}\ \emph {et~al.}(1998)\citenamefont {LeCun},
  \citenamefont {Bottou}, \citenamefont {Bengio},\ and\ \citenamefont
  {Haffner}}]{lecun1998gradient}%
  \BibitemOpen
  \bibfield  {author} {\bibinfo {author} {\bibfnamefont {Y.}~\bibnamefont
  {LeCun}}, \bibinfo {author} {\bibfnamefont {L.}~\bibnamefont {Bottou}},
  \bibinfo {author} {\bibfnamefont {Y.}~\bibnamefont {Bengio}},\ and\ \bibinfo
  {author} {\bibfnamefont {P.}~\bibnamefont {Haffner}},\ }\bibfield  {title}
  {\bibinfo {title} {Gradient-based learning applied to document recognition},\
  }\href@noop {} {\bibfield  {journal} {\bibinfo  {journal} {Proceedings of the
  IEEE}\ }\textbf {\bibinfo {volume} {86}},\ \bibinfo {pages} {2278} (\bibinfo
  {year} {1998})}\BibitemShut {NoStop}%
\bibitem [{\citenamefont {Albawi}\ \emph {et~al.}(2017)\citenamefont {Albawi},
  \citenamefont {Mohammed},\ and\ \citenamefont
  {Al-Zawi}}]{albawi2017understanding}%
  \BibitemOpen
  \bibfield  {author} {\bibinfo {author} {\bibfnamefont {S.}~\bibnamefont
  {Albawi}}, \bibinfo {author} {\bibfnamefont {T.~A.}\ \bibnamefont
  {Mohammed}},\ and\ \bibinfo {author} {\bibfnamefont {S.}~\bibnamefont
  {Al-Zawi}},\ }\bibfield  {title} {\bibinfo {title} {Understanding of a
  convolutional neural network},\ }in\ \href@noop {} {\emph {\bibinfo
  {booktitle} {2017 International Conference on Engineering and Technology
  (ICET)}}}\ (\bibinfo {organization} {Ieee},\ \bibinfo {year} {2017})\ pp.\
  \bibinfo {pages} {1--6}\BibitemShut {NoStop}%
\bibitem [{\citenamefont {Schneeloch}\ \emph {et~al.}(2020)\citenamefont
  {Schneeloch}, \citenamefont {Tison}, \citenamefont {Fanto}, \citenamefont
  {Ray},\ and\ \citenamefont {Alsing}}]{schneeloch2020quantifying}%
  \BibitemOpen
  \bibfield  {author} {\bibinfo {author} {\bibfnamefont {J.}~\bibnamefont
  {Schneeloch}}, \bibinfo {author} {\bibfnamefont {C.~C.}\ \bibnamefont
  {Tison}}, \bibinfo {author} {\bibfnamefont {M.~L.}\ \bibnamefont {Fanto}},
  \bibinfo {author} {\bibfnamefont {S.}~\bibnamefont {Ray}},\ and\ \bibinfo
  {author} {\bibfnamefont {P.~M.}\ \bibnamefont {Alsing}},\ }\bibfield  {title}
  {\bibinfo {title} {Quantifying tripartite entanglement with entropic
  correlations},\ }\href@noop {} {\bibfield  {journal} {\bibinfo  {journal}
  {Physical Review Research}\ }\textbf {\bibinfo {volume} {2}},\ \bibinfo
  {pages} {043152} (\bibinfo {year} {2020})}\BibitemShut {NoStop}%
\bibitem [{\citenamefont {Horn}\ and\ \citenamefont
  {Johnson}(2012)}]{horn2012matrix}%
  \BibitemOpen
  \bibfield  {author} {\bibinfo {author} {\bibfnamefont {R.~A.}\ \bibnamefont
  {Horn}}\ and\ \bibinfo {author} {\bibfnamefont {C.~R.}\ \bibnamefont
  {Johnson}},\ }\href@noop {} {\emph {\bibinfo {title} {Matrix analysis}}}\
  (\bibinfo  {publisher} {Cambridge university press},\ \bibinfo {year}
  {2012})\BibitemShut {NoStop}%
\bibitem [{\citenamefont {Shen}\ and\ \citenamefont
  {Chen}(2020)}]{shen2020construction}%
  \BibitemOpen
  \bibfield  {author} {\bibinfo {author} {\bibfnamefont {Y.}~\bibnamefont
  {Shen}}\ and\ \bibinfo {author} {\bibfnamefont {L.}~\bibnamefont {Chen}},\
  }\bibfield  {title} {\bibinfo {title} {Construction of genuine multipartite
  entangled states},\ }\href@noop {} {\bibfield  {journal} {\bibinfo  {journal}
  {Journal of Physics A: Mathematical and Theoretical}\ }\textbf {\bibinfo
  {volume} {53}},\ \bibinfo {pages} {125302} (\bibinfo {year}
  {2020})}\BibitemShut {NoStop}%
\bibitem [{\citenamefont {Hein}\ \emph {et~al.}(2004)\citenamefont {Hein},
  \citenamefont {Eisert},\ and\ \citenamefont {Briegel}}]{hein2004multiparty}%
  \BibitemOpen
  \bibfield  {author} {\bibinfo {author} {\bibfnamefont {M.}~\bibnamefont
  {Hein}}, \bibinfo {author} {\bibfnamefont {J.}~\bibnamefont {Eisert}},\ and\
  \bibinfo {author} {\bibfnamefont {H.~J.}\ \bibnamefont {Briegel}},\
  }\bibfield  {title} {\bibinfo {title} {Multiparty entanglement in graph
  states},\ }\href@noop {} {\bibfield  {journal} {\bibinfo  {journal} {Physical
  Review A}\ }\textbf {\bibinfo {volume} {69}},\ \bibinfo {pages} {062311}
  (\bibinfo {year} {2004})}\BibitemShut {NoStop}%
\bibitem [{\citenamefont {Hein}\ \emph {et~al.}(2006)\citenamefont {Hein},
  \citenamefont {D{\"u}r}, \citenamefont {Eisert}, \citenamefont {Raussendorf},
  \citenamefont {Nest},\ and\ \citenamefont {Briegel}}]{hein2006entanglement}%
  \BibitemOpen
  \bibfield  {author} {\bibinfo {author} {\bibfnamefont {M.}~\bibnamefont
  {Hein}}, \bibinfo {author} {\bibfnamefont {W.}~\bibnamefont {D{\"u}r}},
  \bibinfo {author} {\bibfnamefont {J.}~\bibnamefont {Eisert}}, \bibinfo
  {author} {\bibfnamefont {R.}~\bibnamefont {Raussendorf}}, \bibinfo {author}
  {\bibfnamefont {M.}~\bibnamefont {Nest}},\ and\ \bibinfo {author}
  {\bibfnamefont {H.-J.}\ \bibnamefont {Briegel}},\ }\bibfield  {title}
  {\bibinfo {title} {Entanglement in graph states and its applications},\
  }\href@noop {} {\bibfield  {journal} {\bibinfo  {journal} {arXiv preprint
  quant-ph/0602096}\ } (\bibinfo {year} {2006})}\BibitemShut {NoStop}%
\bibitem [{\citenamefont {Keet}\ \emph {et~al.}(2010)\citenamefont {Keet},
  \citenamefont {Fortescue}, \citenamefont {Markham},\ and\ \citenamefont
  {Sanders}}]{keet2010quantum}%
  \BibitemOpen
  \bibfield  {author} {\bibinfo {author} {\bibfnamefont {A.}~\bibnamefont
  {Keet}}, \bibinfo {author} {\bibfnamefont {B.}~\bibnamefont {Fortescue}},
  \bibinfo {author} {\bibfnamefont {D.}~\bibnamefont {Markham}},\ and\ \bibinfo
  {author} {\bibfnamefont {B.~C.}\ \bibnamefont {Sanders}},\ }\bibfield
  {title} {\bibinfo {title} {Quantum secret sharing with qudit graph states},\
  }\href@noop {} {\bibfield  {journal} {\bibinfo  {journal} {Physical Review
  A}\ }\textbf {\bibinfo {volume} {82}},\ \bibinfo {pages} {062315} (\bibinfo
  {year} {2010})}\BibitemShut {NoStop}%
\bibitem [{\citenamefont {Looi}\ \emph {et~al.}(2008)\citenamefont {Looi},
  \citenamefont {Yu}, \citenamefont {Gheorghiu},\ and\ \citenamefont
  {Griffiths}}]{looi2008quantum}%
  \BibitemOpen
  \bibfield  {author} {\bibinfo {author} {\bibfnamefont {S.~Y.}\ \bibnamefont
  {Looi}}, \bibinfo {author} {\bibfnamefont {L.}~\bibnamefont {Yu}}, \bibinfo
  {author} {\bibfnamefont {V.}~\bibnamefont {Gheorghiu}},\ and\ \bibinfo
  {author} {\bibfnamefont {R.~B.}\ \bibnamefont {Griffiths}},\ }\bibfield
  {title} {\bibinfo {title} {Quantum-error-correcting codes using qudit graph
  states},\ }\href@noop {} {\bibfield  {journal} {\bibinfo  {journal} {Physical
  Review A}\ }\textbf {\bibinfo {volume} {78}},\ \bibinfo {pages} {042303}
  (\bibinfo {year} {2008})}\BibitemShut {NoStop}%
\bibitem [{\citenamefont {West}\ \emph {et~al.}(2001)\citenamefont {West} \emph
  {et~al.}}]{west2001introduction}%
  \BibitemOpen
  \bibfield  {author} {\bibinfo {author} {\bibfnamefont {D.~B.}\ \bibnamefont
  {West}} \emph {et~al.},\ }\href@noop {} {\emph {\bibinfo {title}
  {Introduction to graph theory}}},\ Vol.~\bibinfo {volume} {2}\ (\bibinfo
  {publisher} {Prentice hall Upper Saddle River},\ \bibinfo {year}
  {2001})\BibitemShut {NoStop}%
\bibitem [{\citenamefont {Lu}\ \emph {et~al.}(2018{\natexlab{b}})\citenamefont
  {Lu}, \citenamefont {Zhao}, \citenamefont {Li}, \citenamefont {Yin},
  \citenamefont {Yuan}, \citenamefont {Hung}, \citenamefont {Chen},
  \citenamefont {Li}, \citenamefont {Liu}, \citenamefont {Peng} \emph
  {et~al.}}]{lu2018entanglement}%
  \BibitemOpen
  \bibfield  {author} {\bibinfo {author} {\bibfnamefont {H.}~\bibnamefont
  {Lu}}, \bibinfo {author} {\bibfnamefont {Q.}~\bibnamefont {Zhao}}, \bibinfo
  {author} {\bibfnamefont {Z.-D.}\ \bibnamefont {Li}}, \bibinfo {author}
  {\bibfnamefont {X.-F.}\ \bibnamefont {Yin}}, \bibinfo {author} {\bibfnamefont
  {X.}~\bibnamefont {Yuan}}, \bibinfo {author} {\bibfnamefont {J.-C.}\
  \bibnamefont {Hung}}, \bibinfo {author} {\bibfnamefont {L.-K.}\ \bibnamefont
  {Chen}}, \bibinfo {author} {\bibfnamefont {L.}~\bibnamefont {Li}}, \bibinfo
  {author} {\bibfnamefont {N.-L.}\ \bibnamefont {Liu}}, \bibinfo {author}
  {\bibfnamefont {C.-Z.}\ \bibnamefont {Peng}}, \emph {et~al.},\ }\bibfield
  {title} {\bibinfo {title} {Entanglement structure: entanglement partitioning
  in multipartite systems and its experimental detection using optimizable
  witnesses},\ }\href@noop {} {\bibfield  {journal} {\bibinfo  {journal}
  {Physical Review X}\ }\textbf {\bibinfo {volume} {8}},\ \bibinfo {pages}
  {021072} (\bibinfo {year} {2018}{\natexlab{b}})}\BibitemShut {NoStop}%
\bibitem [{\citenamefont {Huang}\ \emph {et~al.}(2020)\citenamefont {Huang},
  \citenamefont {Kueng},\ and\ \citenamefont {Preskill}}]{huang2020predicting}%
  \BibitemOpen
  \bibfield  {author} {\bibinfo {author} {\bibfnamefont {H.-Y.}\ \bibnamefont
  {Huang}}, \bibinfo {author} {\bibfnamefont {R.}~\bibnamefont {Kueng}},\ and\
  \bibinfo {author} {\bibfnamefont {J.}~\bibnamefont {Preskill}},\ }\bibfield
  {title} {\bibinfo {title} {Predicting many properties of a quantum system
  from very few measurements},\ }\href@noop {} {\bibfield  {journal} {\bibinfo
  {journal} {Nature Physics}\ }\textbf {\bibinfo {volume} {16}},\ \bibinfo
  {pages} {1050} (\bibinfo {year} {2020})}\BibitemShut {NoStop}%
\end{thebibliography}%

\end{document}